 \documentclass[final,5p,times,twocolumn,compress]{elsarticle}

\usepackage{amssymb}
\usepackage{amsmath}
\usepackage{lipsum}
\usepackage{color}

\journal{Physics Letters A}

\begin{document}

\begin{frontmatter}

\title{Quantum recharging by shortcut to adiabaticity}

\author[first]{Shi-fan Qi}
\affiliation[first]{organization={College of Physics and Hebei Key Laboratory of Photophysics Research and Application, Hebei Normal University},
            city={Shijiazhuang},
            postcode={050024}, 
            country={China}}
\author[second]{Jun Jing}          
\affiliation[second]{organization={School of Physics, Zhejiang University},
            city={Hangzhou},
            postcode={310027}, 
            country={China}}
\ead{jingjun@zju.edu.cn}
            
\begin{abstract}
Quantum battery concerns about population redistribution and energy dispatch over controllable quantum systems. Under unitary transformation, ergotropy rather than energy plays an essential role in describing the accumulated useful work. Thus, the charging and recharging of quantum batteries are distinct from the electric-energy input and reuse of classical batteries. In this work, we focus on recharging a three-level quantum battery that has been exhausted under self-discharging and work extraction. We find that the quantum battery cannot be fully refreshed with the maximum ergotropy only by the driving pulses for unitary charging. For an efficient refreshment of the quantum battery, we propose a fast and stable recharging protocol based on postselection and shortcut to adiabaticity. More than accelerating the adiabatic passage for charging, the protocol can eliminate unextractable energy and is robust against driving errors and environmental decoherence. Our protocol is energy-saving and experimental-feasible, even in systems with the forbidden transition.
\end{abstract}

\begin{keyword}
Quantum battery \sep Shortcut to adiabaticity \sep Quantum optics \sep Systematic errors \sep Environmental decoherence
\end{keyword}

\end{frontmatter}

\section{Introduction}
\label{introduction}
Recent advances in quantum thermodynamics~\cite{thermo1,thermo2} have stimulated the conceptual generalization about the maximal capacity of an interested system to transfer between a passive state and an active state. Alicki and Fannes pioneered a quantum device termed quantum battery (QB) that can store and release energy under unitary transformation in a controllable manner to mimic its counterpart in the classical world~\cite{workextract}. In exploiting its potential advantages over the classical battery, many careful investigations~\cite{batquancell,batenhanc,batchargemediateenergy,batspinchain,batmanybody,batopensys,batmantbody2,
batultrafast,batfluc,boundscapacity,batdark,qdsyk,batennoise,batspeedup,batmagnon,Optimalcontrol,batmediator,batvacuum,batmeasure,batcoll,batcapa,batdipole,twophotoncharging,Frustrating,threelevelDicke,Nonreciprocal,nonHermitian} have been carried out, targeting faster charging rate, more extractable energy, and higher stability in control.

A quantum battery~\cite{colloquiumqb} can be charged either by a classical driving~\cite{stablebattery,classical,batteryexp} or by the interaction with an energy-filled auxiliary system (quantum charger)~\cite{batspeedup,batmagnon,batmediator,batlarge,batrepeat}. Conventional studies were initiated primarily around promoting and optimizing the charging performance in quantum regimes. To name a few, how can the presence of quantum coherence or entanglement affect the energy storage~\cite{entcohqb,demonqb}, how to simultaneously achieve a full charged state and reduce the charging period~\cite{batenhanc,batultrafast,highpower}, and how to realize a stable charging with no energy backflow after the charging is completed. Besides ergotropy~\cite{Workfluctuation} (the energy that can be extracted by unitary transformation for work) and charging power, stable charging was another important measure in quantum charging~\cite{stablebattery,stable2}, which avoids the extremely precise control over a simple $\pi$ pulse or Rabi oscillation~\cite{stablebattery,stable2,stable3}. However, few existing works are concerned about a renewable QB, with respect to the self-discharging process and energy extraction. Recharging is one of the bottlenecks in preventing the widespread use of quantum batteries.

In this work, we propose a recharging protocol for a three-level QB, using a shortcut to adiabaticity (STA) technique~\cite{STAcd} and state postselection. Both of them contribute to the toolbox of quantum control, enabling highly efficient dynamical operations in modern quantum technologies. For STA, we here employ the counterdiabatic (CD) driving method~\cite{stacd2}, also named quantum transitionless driving~\cite{Berry2009}. In general, a CD Hamiltonian can be constructed as~\cite{STAcd,stacd2,Berry2009,staatom,staatom2,staberry}
\begin{equation}\label{Hcdgeneral}
H_{\rm CD}=i\sum_n\left[1-|n(t)\rangle\langle n(t)|\right]|\dot{n}(t)\rangle\langle n(t)|,
\end{equation}
where $|n(t)\rangle$ is the instantaneous eigenvectors of the original time-dependent Hamiltonian $H(t)$ and $|\dot{n}(t)\rangle$ means its time derivative. The charging protocol aided by the CD driving can move the battery system exactly along the adiabatic path at a much faster speed than those based on the stimulated Raman adiabatic passage (STIRAP)~\cite{stablebattery,batteryexp}. However, when the battery starts from a passive state with finite energy yet vanishing ergotropy, rather than the ground state as commonly considered in literature~\cite{stablebattery,batteryexp}, it can not be fully recharged with the maximum ergotropy by any unitary transformation including the STA evolution. We find that this problem can be dealt with by a postselection method with a considerable success probability.

The rest of this work is organized as follows. In Sec.~\ref{Sec:quantum}, we briefly recall the basic concepts of QB. After presenting the evolution process when the battery is subject to self-discharging (caused by the presence of environment) and work extraction, we illustrate our recharging protocol on postselection (projective measurement) and counterdiabatic driving for the three-level QB in a cascade type. It is shown by the numerical simulation of population and ergotropy that our protocol can restore the battery to the most active state with the maximum ergotropy. We estimate the robustness of our recharging protocol against the systematic errors arising from the driving pulses in Sec.~\ref{Sec:drivingerror} and the environmental noises in Sec.~\ref{Sec:decoherence}, respectively. In Sec.~\ref{Sec:physic}, we show that in the superconducting qutrit systems, the CD Hamiltonian can be achieved by a two-photon process to avoid the forbidden transition. In Sec.~\ref{Sec:ener}, we discuss the energetic costs of STA control and projective measurement. The conclusion is provided in Sec.~\ref{Sec:conclusion}. In~\ref{appa}, we compare the charging process with the conventional STIRAP and the STA protocols.

\section{Quantum recharging protocol}\label{Sec:quantum}
\begin{figure}[htbp]
\centering
\includegraphics[width=0.45\textwidth]{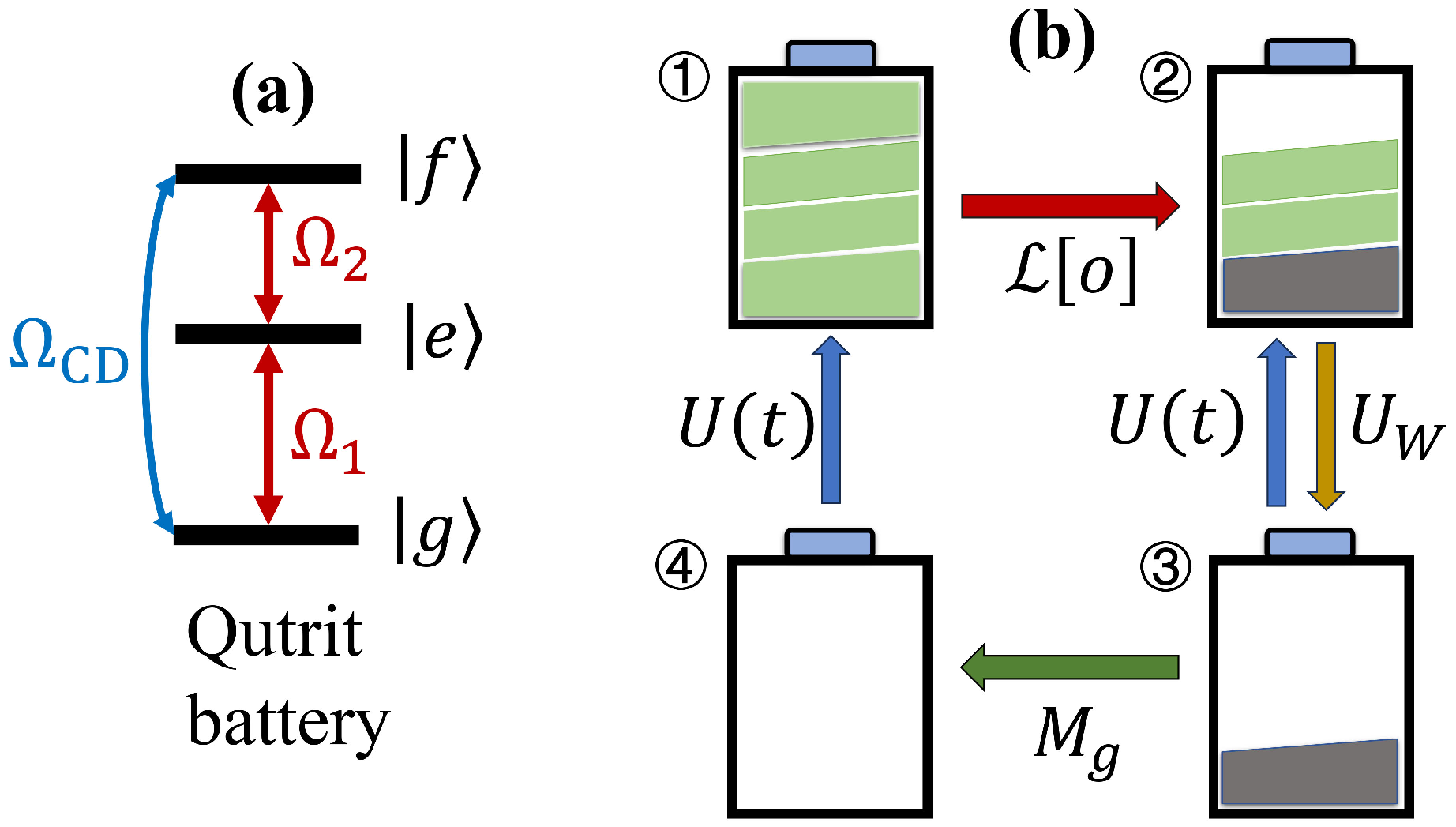}
\caption{(a) Diagram of a three-level QB of the cascade type under resonant driving pulses. The transition between the ground state $|g\rangle$ and the intermediate state $|e\rangle$ and that between $|e\rangle$ and the excited state $|f\rangle$ are coupled to the driving pulses with Rabi frequency $\Omega_1$ and $\Omega_2$, respectively. The ancillary driving pulse $\Omega_{\rm CD}$ is applied to the transition $|g\rangle\leftrightarrow|f\rangle$. (b) Diagram of our recharging protocol, including $\textcircled{\scriptsize{1}}\rightarrow\textcircled{\scriptsize{2}}$: QB self-discharging induced by decoherence $\mathcal{L}[o]$, $\textcircled{\scriptsize{2}}\leftrightarrow\textcircled{\scriptsize{3}}$: work extraction by unitary transformation $U_W$ and recharging operation $U(t)$ assisted by STA, and $\textcircled{\scriptsize{3}}\rightarrow\textcircled{\scriptsize{4}}\rightarrow\textcircled{\scriptsize{1}}$: postselection by the projective measurement $M_g$ and recharging $U(t)$ assisted by STA. The battery energy is divided into extractable (green) and unextractable (gray) parts.}\label{diagram}
\end{figure}
A non-degenerate $n$-level QB can be described by the Hamiltonian
\begin{equation}\label{nondeHam}
H_0=\sum^n_{j=1}\epsilon_j|\epsilon_j\rangle\langle\epsilon_j|,
\end{equation}
where $\epsilon_j$'s are the eigen-energies of the bare system ordered by $\epsilon_1<\epsilon_2<\cdots<\epsilon_n$. The internal energy of such a QB is given by ${\rm Tr}[\rho H_0]$, where $\rho$ is the density matrix. A QB is on charging such that the internal energy increases when its state varies from $\rho$ to $\rho'$, i.e., ${\rm Tr}[(\rho'-\rho)H_0]\ge0$. The opposite variation can be regarded as discharging.

Ergotropy is the central quantity in the study of QB, which is defined as the maximum amount of available work that can be extracted from the battery through unitary transformation~\cite{workextract,batteryexp}. It is given by
\begin{equation}\label{ergoint}
\xi(t)={\rm Tr}[\rho(t)H_0]-\underset{U_{\rm w}\in\mathcal{U}}{\rm min}\left\{{\rm Tr}[U_{\rm w}\rho(t)U_{\rm w}^\dag H_0]\right\},
\end{equation}
where the minimization is taken over the set $\mathcal{U}$ of the unitary operators $U_{\rm w}$ acting on the system. The most successful energy-extraction operation can transform the QB system to a passive state~\cite{workextract}. Given the system density matrix $\rho$, there is a unique passive state minimizing ${\rm Tr}[U_{\rm w}\rho(t)U_{\rm w}^\dag H_0]$. With the spectral decomposition of the battery state $\rho=\sum^n_{k=1}p_k|p_k\rangle\langle p_k|$, $p_1\ge p_2\ge\cdots\ge p_n$, the ergotropy can be written as
\begin{equation}\label{ergoint2}
\xi(t)=\sum^n_{k,j=1}p_k\epsilon_j\left(|\langle p_k|\epsilon_j\rangle|^2-\delta_{kj}\right),
\end{equation}
where $\delta_{kj}$ denotes the Kronecker delta function. Ergotropy rather than energy evaluates the performance of a QB under discharging and recharging.

The QB system in this work is a cascade-type three-level qutrit as shown in Fig.~\ref{diagram}(a). The ground state, the intermediate state, and the excited state are labeled with $|g\rangle$, $|e\rangle$, and $|f\rangle$, respectively. The bare Hamiltonian for QB can be written as ($\hbar\equiv1$)
\begin{equation}\label{Hambare}
H_0=\omega_e|e\rangle\langle e|+\omega_f|f\rangle\langle f|,
\end{equation}
where the ground-state energy is set as $\omega_g\equiv0$ with no loss of generality. During the charging process, two microwave fields with Rabi frequencies $\Omega_1$ and $\Omega_2$ are resonantly coupled to the $|g\rangle\leftrightarrow|e\rangle$ and $|e\rangle\leftrightarrow|f\rangle$ transitions, respectively. And the ancillary pulse $\Omega_{\rm CD}$ for STA represents the counterdiabatic driving applied to the $|g\rangle\leftrightarrow|f\rangle$ transition.

Figure~\ref{diagram}(b) is a flow diagram for our recharging protocol. On stage $\textcircled{\scriptsize{1}}$, the QB starts from a full-charged state. It cannot be an ideally isolated system and will be spontaneously self-discharged in the presence of an environment. As described by a Lindblad dissipator $\mathcal{L}[o]$, gradually the QB becomes a less active state on stage $\textcircled{\scriptsize{2}}$ besides losing energy. In other words, the QB energy on stage $\textcircled{\scriptsize{2}}$ cannot be fully extracted. The extractable and unextractable energies are indicated by the green and gray colors, respectively. After the work extraction performed by the unitary transformation $U_{\rm W}$, the QB becomes a passive state on stage $\textcircled{\scriptsize{3}}$, which is the initial state for the following recharging process. The detailed descriptions of self-discharging and work extraction are provided in Sec.~\ref{Sec:self-discharging}. On stage $\textcircled{\scriptsize{3}}$, one has two choices for recharging. One can directly apply the STA driving pulses in Fig.~\ref{diagram}(a) to the QB, which is denoted with $U(t)$. The optimal result one can obtain is to restore the QB to the partial active state $\textcircled{\scriptsize{2}}$ before the work extraction. Alternatively, one can use the postselection performed by the projective measurement on the ground level $|g\rangle$ to transform the QB to an empty state on stage $\textcircled{\scriptsize{4}}$ and then realize the full charging via the STA evolution. The details are presented in Sec.~\ref{Sec:charging}.

\subsection{Self-discharging and work extraction}\label{Sec:self-discharging}
The self-discharging dynamics of the QB as $\textcircled{\scriptsize{1}}\rightarrow\textcircled{\scriptsize{2}}$ shown in Fig.~\ref{diagram}(b) is governed by the Lindblad master equation,
\begin{equation}\label{lindblad}
\frac{\partial{\rho}}{\partial t}=-i[H_0,\rho]+\frac{1}{2}\sum_{n\in\{e,f\}}\left(\gamma_n\mathcal{L}[\sigma^-_n]+
\gamma^z_n\mathcal{L}[\sigma^z_n]\right),
\end{equation}
where the super-operation $\mathcal{L}[o]$ is defined as
\begin{equation}\label{superoperator}
\mathcal{L}[o]\equiv2o\rho o^\dag-o^\dag o\rho-\rho o^\dag o
\end{equation}
with the system operator $o$. Here $\rho$ is density matrix of the three-level QB, $\sigma_e^-=|g\rangle\langle e|$, $\sigma_f^-=|e\rangle\langle f|$, $\sigma^z_e=|e\rangle\langle e|-|g\rangle\langle g|$, and $\sigma^z_f=|f\rangle\langle f|-|e\rangle\langle e|$. $\gamma_n$ and $\gamma^z_n$, $n\in\{e,f\}$, are respectively the decay and dephasing rates. In recent superconducting qutrit systems~\cite{batteryexp,measurecost2}, the decay rates satisfy $\gamma_f>\gamma_e$ and are of the same order of magnitude as the dephasing rates. For simplicity and without loss of generality, we assume $\gamma_f=1.5\gamma_e$ and $\gamma^z_f=\gamma^z_e=2\gamma_e$ in the following numerical results.

In the space spanned by $\{|g\rangle, |e\rangle, |f\rangle\}$, the master equation in Eq.~(\ref{lindblad}) can be resolved into the time evolution of the diagonal elements
\begin{equation}\label{diagnoal}
\begin{aligned}
\frac{\partial\rho_{ff}}{\partial t}&=-\gamma_f\rho_{ff},\quad \frac{\partial\rho_{ee}}{\partial t}=\gamma_f\rho_{ff}-\gamma_e\rho_{ee},\\
\frac{\partial\rho_{gg}}{\partial t}&=\gamma_e\rho_{ee},
\end{aligned}
\end{equation}
and that of the off-diagonal elements
\begin{equation}\label{nondiagnoal}
\begin{aligned}
\frac{\partial\rho_{fe}}{\partial t}&=-i(\omega_f-\omega_e)\rho_{fe}-\frac{4\gamma^z_f+\gamma^z_e+\gamma_f+\gamma_e}{2}\rho_{fe},\\
\frac{\partial\rho_{fg}}{\partial t}&=-i\omega_f\rho_{fg}-\frac{\gamma^z_f+\gamma^z_e+\gamma_f}{2}\rho_{fg},\\
\frac{\partial\rho_{eg}}{\partial t}&=-i\omega_e\rho_{eg}-\frac{\gamma^z_f+4\gamma^z_e+\gamma_e}{2}\rho_{eg}.
\end{aligned}
\end{equation}

\begin{figure}[htbp]
\centering
\includegraphics[width=0.45\textwidth]{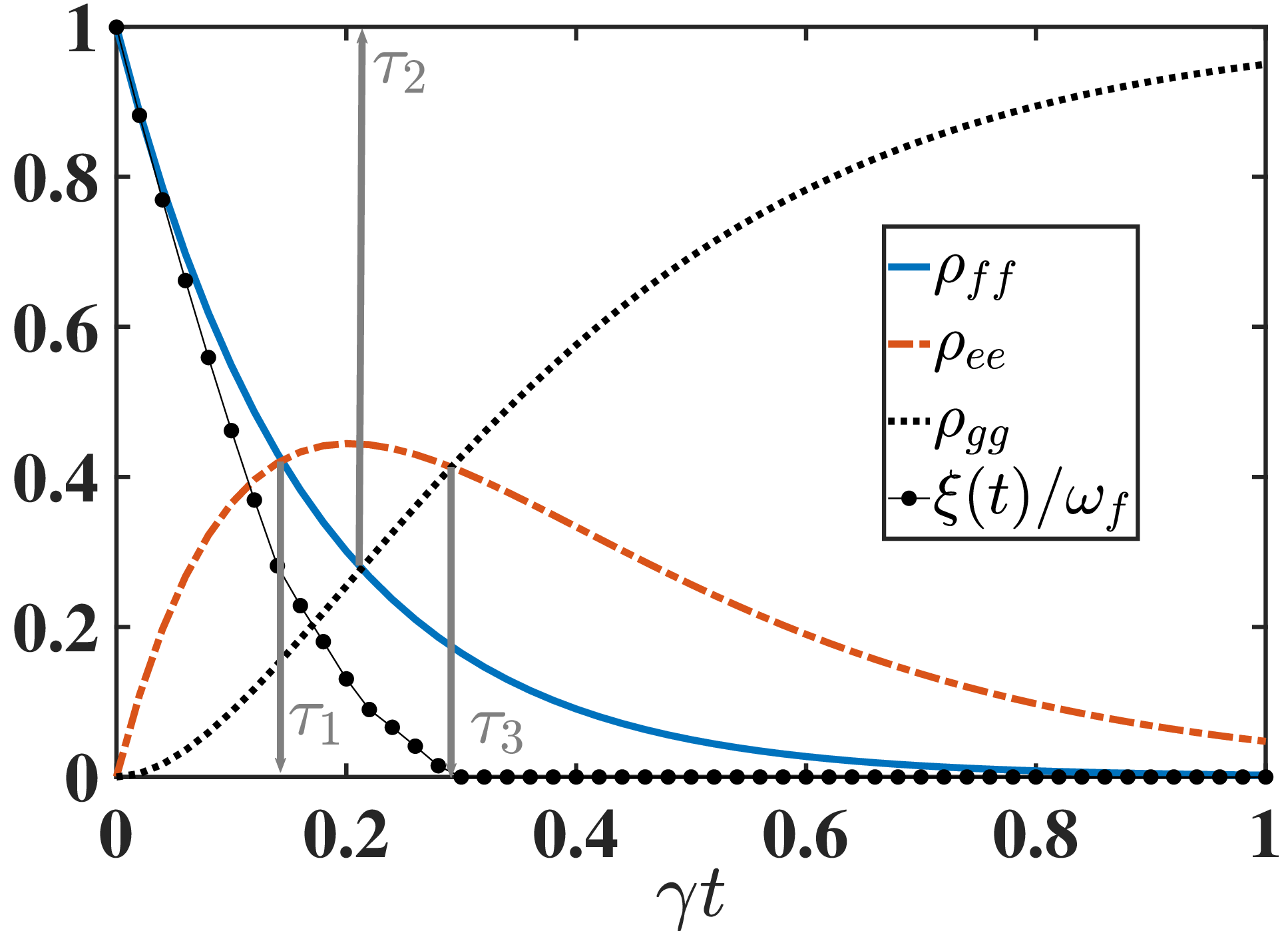}
\caption{Populations on the three levels and battery ergotropy during the self-discharging process. The decoherence rates are set as $\gamma_e=\gamma$, $\gamma_f=1.5\gamma$, and $\gamma^z_f=\gamma^z_e=2\gamma$. The transition frequencies are fixed as $\omega_e=10^5\gamma$ and $\omega_f=1.7\times 10^5\gamma$. }\label{selfdischarge}
\end{figure}
As stage $\textcircled{\scriptsize{1}}$ shown in Fig.~\ref{diagram}(b), the QB is supposed to be fully charged at the initial time, i.e., $\rho_{ff}(0)=1$. By Eqs.~(\ref{diagnoal}) and (\ref{nondiagnoal}), we have
\begin{equation}\label{element}
\begin{aligned}
\rho_{ff}(t)&=e^{-\gamma_ft},\\
\rho_{ee}(t)&=\frac{\gamma_f}{\gamma_e-\gamma_f}\left(e^{-\gamma_ft}-e^{-\gamma_et}\right),\\
\rho_{gg}(t)&=1-\frac{\gamma_ee^{-\gamma_ft}-\gamma_fe^{-\gamma_et}}{\gamma_e-\gamma_f},\\
\rho_{fe}(t)&=\rho_{fg}(t)=\rho_{eg}(t)=0.
\end{aligned}
\end{equation}
Normally we have therefore three crossing moments $\tau_j$, $j=1,2,3$, to have $\rho_{ff}(\tau_1)=\rho_{ee}(\tau_1)$, $\rho_{ff}(\tau_2)=\rho_{gg}(\tau_2)$, and $\rho_{ee}(\tau_3)=\rho_{gg}(\tau_3)$ during the self-discharging described by Fig.~\ref{selfdischarge}. In particular, we have
\begin{equation}\label{tau1}
\tau_1=\frac{\ln(2\gamma_f-\gamma_e)-\ln\gamma_f}{\gamma_f-\gamma_e}.
\end{equation}

According to the definition in Eq.~(\ref{ergoint2}), the time-evolved ergotropy for various situations can be written as
\begin{equation}\label{ergotro}
\xi=\left\{
\begin{aligned}
&\omega_f(\rho_{ff}-\rho_{gg}),&\rho_{ff}\ge\rho_{ee}\ge\rho_{gg}\\
&\omega_f(\rho_{ff}-\rho_{gg})+\omega_e(\rho_{ee}-\rho_{ff}),&\rho_{ee}\ge\rho_{ff}\ge\rho_{gg}\\
&\omega_e(\rho_{ee}-\rho_{gg}),&\rho_{ee}\ge\rho_{gg}\ge\rho_{ff}\\
&0,&\rho_{gg}\ge\rho_{ee}\ge\rho_{ff}
\end{aligned}
\right.
\end{equation}
where for brevity we have dropped the explicit time dependence. During the interval $\tau\in[0, \tau_1]$ when the populations satisfy $\rho_{ff}(\tau)\ge\rho_{ee}(\tau)>\rho_{gg}(\tau)$, we have
\begin{equation}\label{ergo}
\xi(\tau)=\omega_f[\rho_{ff}(\tau)-\rho_{gg}(\tau)].
\end{equation}
After $\tau_3$, the QB becomes completely passive when $\rho_{gg}>\rho_{ee}>\rho_{ff}$, i.e., no energy can be extracted for work from the battery with unitary transformation. Yet we can focus merely on the interval $0\leq t<\tau$ with $\tau\leq\tau_1$ since the QB ergotropy has become sufficiently low around $\tau_1$. Then on stage $\textcircled{\scriptsize{2}}$, one can start a work-extraction process, i.e., $\textcircled{\scriptsize{2}}\rightarrow\textcircled{\scriptsize{3}}$ in Fig.~\ref{diagram}(b), on the QB. Note $\tau_1$ has been determined by Eq.~(\ref{tau1}) in advance, so that both work-extraction and the following recharging can be performed on any state $\rho(\tau<\tau_1)$. By Eq.~(\ref{ergoint2}), one can find that the work extraction yields the population swapping between levels $|g\rangle$ and $|f\rangle$ while the population on $|e\rangle$ remains invariant. The extraction operation can thus be physically realized by the following unitary transformation as
\begin{equation}\label{Uwork}
\begin{aligned}
U_{\rm w}&=\begin{bmatrix}
0 & 0 & 1\\
0 & 1 & 0\\
1 & 0 & 0
\end{bmatrix},
\end{aligned}
\end{equation}
up to the local phases. In fact, any unitary operation that swaps the populations on $|g\rangle$ and $|f\rangle$ without extra effects is theoretically feasible, by which the QB density matrix turns out to be $U_{\rm w}\rho(\tau)U^\dag_{\rm w}$. In comparison to the discharging dynamics, the duration of the work-extracting operation $U_{\rm w}$ can be omitted.

\subsection{Recharging by shortcut to adiabaticity}\label{Sec:charging}
We present in this section our recharging protocol assisted by counterdiabatic driving. It starts after the self-discharging process lasting a period of $\tau<\tau_1$ and the instantaneous work extraction. The initial state for the QB recharging process is written as $\tilde{\rho}(0)=U_{\rm w}\rho(\tau)U^\dag_{\rm w}$. Thus by Eqs.~(\ref{element}) and (\ref{Uwork}), we have
\begin{equation}\label{initialstate}
\begin{aligned}
\tilde{\rho}_{ff}(0)&=\rho_{gg}(\tau)=1-\frac{\gamma_ee^{-\gamma_f\tau}-\gamma_fe^{-\gamma_e\tau}}{\gamma_e-\gamma_f},\\
\tilde{\rho}_{ee}(0)&=\rho_{ee}(\tau)=\frac{\gamma_f}{\gamma_e-\gamma_f}\left(e^{-\gamma_f\tau}-e^{-\gamma_e\tau}\right),\\
\tilde{\rho}_{gg}(0)&=\rho_{ff}(\tau)=e^{-\gamma_f\tau},\\
\tilde{\rho}_{fe}(0)&=\tilde{\rho}_{fg}(0)=\tilde{\rho}_{eg}(0)=0.
\end{aligned}
\end{equation}
Here the tilde superscript distinguishes the starting point on stage $\textcircled{\scriptsize{3}}$ for the recharging process, which is distinct from that on stage $\textcircled{\scriptsize{1}}$. The state in Eq.~(\ref{initialstate}) is an energetic yet passive state since $\tilde{\rho}_{gg}(0)\ge\tilde{\rho}_{ee}(0)\ge\tilde{\rho}_{ff}(0)$. The energy stored in QB is nonzero but is unable to be extracted. The recharging timescale is normally much shorter than the self-charging period and then can be omitted in the ideal situation. We will discuss the nonideal scenario in Sec.~\ref{Sec:decoherence}.

As shown in Fig.~\ref{diagram}(a), the Hamiltonian for the three-level system of QB coupled to the external driving fields reads
\begin{equation}\label{Hamt}
H_{\rm tot}(t)=H_0+V(t),
\end{equation}
where the driving term is
\begin{equation}\label{drivHam}
V(t)=\Omega_1(t)e^{i\omega_1t}|g\rangle\langle e|+\Omega_2(t)e^{i\omega_2t}|e\rangle\langle f|+{\rm H.c.}
\end{equation}
with the Rabi frequencies $\Omega_j$ and the driving frequencies $\omega_j$, $j=1,2$. In the rotating frame with respect to $U_0(t)=\exp(iH_0t)=\exp(i\omega_et|e\rangle\langle e|+i\omega_ft|f\rangle\langle f|)$, the full Hamiltonian in Eq.~(\ref{Hamt}) can be rewritten as
\begin{equation}\label{Ht}
\begin{aligned}
H(t)&=U_0(t)H_{\rm tot}(t)U^\dag_0(t)-iU_0(t)\dot{U}^\dag_0(t)\\
&=\Omega_1(t)|g\rangle\langle e|+\Omega_2(t)|e\rangle\langle f|+{\rm H.c.}.
\end{aligned}
\end{equation}
Here the driving frequencies satisfy the one-photon resonant condition, i.e., $\omega_1=\omega_e$ and $\omega_2=\omega_f-\omega_e$. The eigenvectors of the Hamiltonian in Eq.~(\ref{Ht}) are
\begin{equation}\label{eigenstr}
\begin{aligned}
&|\lambda_0(t)\rangle=\cos\theta(t)|g\rangle-\sin\theta(t)|f\rangle,\\
&|\lambda_{\pm}(t)\rangle=[\sin\theta(t)|g\rangle\pm|e\rangle+\cos\theta(t)|f\rangle]/\sqrt{2},
\end{aligned}
\end{equation}
where $\tan\theta(t)=\Omega_1(t)/\Omega_2(t)$. Their corresponding eigenvalues are $\lambda_0(t)=0$ and $\lambda_{\pm}(t)=\pm\Omega(t)$ with the driving strength $\Omega(t)=\sqrt{\Omega^2_1(t)+\Omega^2_2(t)}$. The boundary conditions of driving pulses are set as $\theta(0)=0$ and $\theta(\tau_c)=\pi/2$, i.e., $\Omega_1(0)=0$, $\Omega_2(0)\ne0$ and $\Omega_1(\tau_c)\ne0$, $\Omega_2(\tau_c)=0$, where $\tau_c$ is the charging period. They are popularly used in both STIRAP~\cite{STIRAP,stablebattery} and STA protocols~\cite{STAcd,Berry2009,staatom} for state transfer.

By virtue of the standard method in Eq.~(\ref{Hcdgeneral}) and the eigen-structure in Eq.~(\ref{eigenstr}), $H_{\rm CD}$ in this work can be obtained as
\begin{equation}\label{Hcd}
H_{\rm CD}(t)=i\Omega_{\rm CD}(t)|g\rangle\langle f|-i\Omega_{\rm CD}(t)|f\rangle\langle g|,
\end{equation}
where
\begin{equation}\label{omegacd}
\Omega_{\rm CD}(t)=\dot{\theta}(t)=\frac{\dot{\Omega}_1(t)\Omega_2(t)-\Omega_1(t)\dot{\Omega}_2(t)}{\Omega^2(t)}.
\end{equation}
Consequently, the STA Hamiltonian is obtained by
\begin{equation}\label{Hsta}
\begin{aligned}
&H_{\rm STA}(t)=H(t)+H_{\rm CD}(t)\\
&=\Omega_1(t)|g\rangle\langle e|+\Omega_2(t)|e\rangle\langle f|+i\Omega_{\rm CD}(t)|g\rangle\langle f|+{\rm H.c.}.
\end{aligned}
\end{equation}
The time-evolution operator $U(t)$ under $H_{\rm STA}$ is then given by
\begin{equation}\label{timeU}
\begin{aligned}
U(t)&=\mathcal{T}_{\leftarrow}\exp\left[-i\int^t_0H_{\rm STA}(t')dt'\right]\\
&=|\lambda_0(t)\rangle\langle\lambda_0(0)|+e^{-i\phi(t)}|\lambda_+(t)\rangle\langle\lambda_+(0)|+e^{i\phi(t)}|\lambda_-(t)\rangle\langle\lambda_-(0)|,
\end{aligned}
\end{equation}
where $\phi(t)\equiv\int^t_0\Omega(t')dt'$. In the space spanned by $\{|g\rangle, |e\rangle, |f\rangle\}$, $U(t)$ can be written as
\begin{equation}\label{timeUspace}
\begin{aligned}
U(t)&=\begin{bmatrix}
\cos\theta(t) & -i\sin\phi(t)\sin\theta(t) & \cos\phi(t)\sin\theta(t)\\
0 & \cos\phi(t) & -i\sin\phi(t)\\
-\sin\theta(t) & -i\sin\phi(t)\cos\theta(t) & \cos\phi(t)\cos\theta(t)
\end{bmatrix},
\end{aligned}
\end{equation}
whose initial condition is consistent with $\theta(0)=0$. Then the transition $\textcircled{\scriptsize{3}}\rightarrow\textcircled{\scriptsize{2}}$ is described by $\tilde{\rho}(t)=U(t)\tilde{\rho}(0)U^\dag(t)$ and through which the ergotropy becomes
\begin{equation}\label{ergotropy}
\begin{aligned}
\xi(t)&={\rm Tr}[\tilde{\rho}(t)H_0]-{\rm Tr}[\tilde{\rho}(0)H_0]\\
&=\omega_e\sin^2\phi(t)\left[\tilde{\rho}_{ff}(0)-\tilde{\rho}_{ee}(0)\right]+\omega_f\sin^2\theta(t)\tilde{\rho}_{gg}(0)\\
&+\omega_f\cos^2\theta(t)[\sin^2\phi(t)\tilde{\rho}_{ee}(0)+\cos^2\phi(t)\tilde{\rho}_{ff}(0)]\\
&-\omega_f\tilde{\rho}_{ff}(0).
\end{aligned}
\end{equation}
At the end of the recharging process, we have
\begin{equation}\label{ergotropyend}
\xi(\tau_c)=\omega_e\sin^2\phi_c[\tilde{\rho}_{ff}(0)-\tilde{\rho}_{ee}(0)]+\omega_f[\tilde{\rho}_{gg}(0)-\tilde{\rho}_{ff}(0)],
\end{equation}
where $\phi_c=\phi(\tau_c)$. Here we have applied the boundary condition $\theta(\tau_c)=\pi/2$. Since $\tilde{\rho}_{ee}(0)>\tilde{\rho}_{ff}(0)$, the maximum value of $\xi(\tau_c)$ can reach
\begin{equation}\label{ergotropymax}
\begin{aligned}
\xi_{\rm max}(\tau_c)&=\omega_f[\tilde{\rho}_{gg}(0)-\tilde{\rho}_{ff}(0)]\\
&=\omega_f[\rho_{ff}(\tau)-\rho_{gg}(\tau)]=\xi(\tau)
\end{aligned}
\end{equation}
when $\phi_c=k\pi$ with $k$ an integer. By reference to Eq.~(\ref{ergo}), the battery recovers the state on stage $\textcircled{\scriptsize{2}}$ before the energy was extracted, i.e., $\tilde{\rho}(\tau_c)=U(\tau_c)\tilde{\rho}(0)U^\dag(\tau_c)=\rho(\tau)$. $\xi_{\rm max}(\tau_c)$ in Eq.~(\ref{ergotropymax}) is the maximal ergotropy of the battery obtained through the STA evolution, which is less than $\omega_f$. It is then found that the battery in a mixed initial state $\tilde{\rho}(0)$ cannot be fully recharged via unitary transformation. Also, the recharging process is unstable since the final state is not an eigenstate of Hamiltonian. With a nonvanishing interaction Hamiltonian $V(t>\tau_c)\ne0$, the ergotropy of QB will decrease and cannot be maintained as $\xi_{\rm max}(\tau_c)$.

To avoid these defects, we can apply a projective measurement as described by $\textcircled{\scriptsize{3}}\rightarrow \textcircled{\scriptsize{4}}$ in Fig.~\ref{diagram}(b) before launching the STA charging protocol. An instantaneous projection $M_g\equiv|g\rangle\langle g|$ on the qutrit battery would transform the density operator $\tilde{\rho}(0)$ to be
\begin{equation}\label{densitymea}
\tilde{\rho}^M(0)=|g\rangle\langle g|
\end{equation}
with a success probability $P_g\equiv\tilde{\rho}_{gg}(0)=\rho_{ff}(\tau)=\exp(-\gamma_f\tau)$ depending on the self-discharging period $\tau$. Evidently a less $\tau$ gives rise to a larger $P_g$. For example, one can observe in Fig.~\ref{selfdischarge} that $P_g$ is over $40\%$ even when $\tau=\tau_1$, which is much greater than the success probability $P_f=\tilde{\rho}_{ff}(0)=\rho_{gg}(\tau)$ for the projection $M_f\equiv|f\rangle\langle f|$.

Under the driving Hamiltonian $H_{\rm STA}$ in Eq.~(\ref{Hsta}), the time-dependent density matrix evolves as
\begin{equation}\label{densityt}
\tilde{\rho}^M(t)=U(t)\tilde{\rho}^M(0)U^\dag(t)=|\lambda_0(t)\rangle\langle\lambda_0(t)|,
\end{equation}
where $|\lambda_0(t)\rangle$ is the dark state in Eq.~(\ref{Ht}). Due to the facts that $\lambda_0(t)=0$ and $\langle\lambda_0(t)|\dot{\lambda}_0(t)\rangle=0$, no quantal phase is accumulated during the evolution. Thus any quantity including the gained ergotropy $\xi(t)$ has no oscillating behavior. At the end of the recharging process, we can have a fully population-inverted state
\begin{equation}\label{densityend}
\tilde{\rho}^M(\tau_c)=|f\rangle\langle f|.
\end{equation}
The battery now returns to stage $\textcircled{\scriptsize{1}}$ in Fig.~\ref{diagram}(b), endowed with a maximum ergotropy $\xi(\tau_c)=\omega_f$. And the recharging is stable without precise control over the charging period $\tau_c$, provided that $\Omega_1(t)\ne0$ and $\Omega_2(t)=0$ when $t\ge\tau_c$. It means that either (1) the QB remains in the fully charged state $|f\rangle$, i.e., the dark state of the full Hamiltonian in Eq.~(\ref{Hamt}) for $t>\tau_c$ with a nonvanishing interaction Hamiltonian $V(t>\tau_c)\ne0$; or (2) the dynamics of the three-level QB is under the bare Hamiltonian $H_0$ in Eq.~(\ref{Hambare}) with $V(t>\tau_c)=0$, and then the fully charged state is still invariant under the ideal situation since it is an eigenstate of $H_0$. Stable charging can also be achieved by conventional protocols based on STIRAP~\cite{stablebattery,batteryexp}, which is however much slower than our STA protocol by using an extra counterdiabatic driving. Details are provided in~\ref{appa}.

\subsection{Numerical simulations of the recharging process}\label{Sec:numer}
\begin{figure}[htbp]
\centering
\includegraphics[width=0.45\textwidth,height=0.32\textwidth]{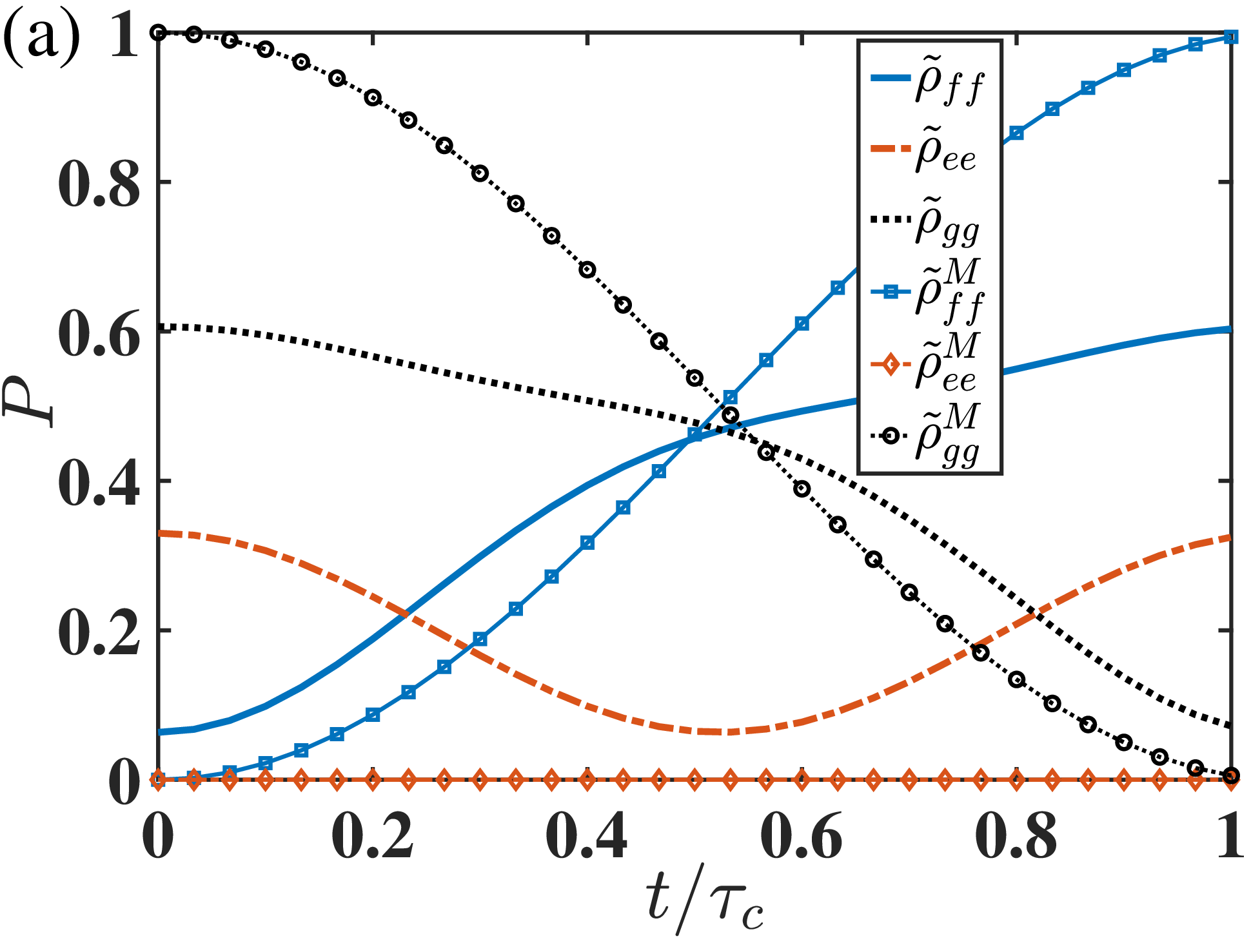}
\includegraphics[width=0.45\textwidth,height=0.32\textwidth]{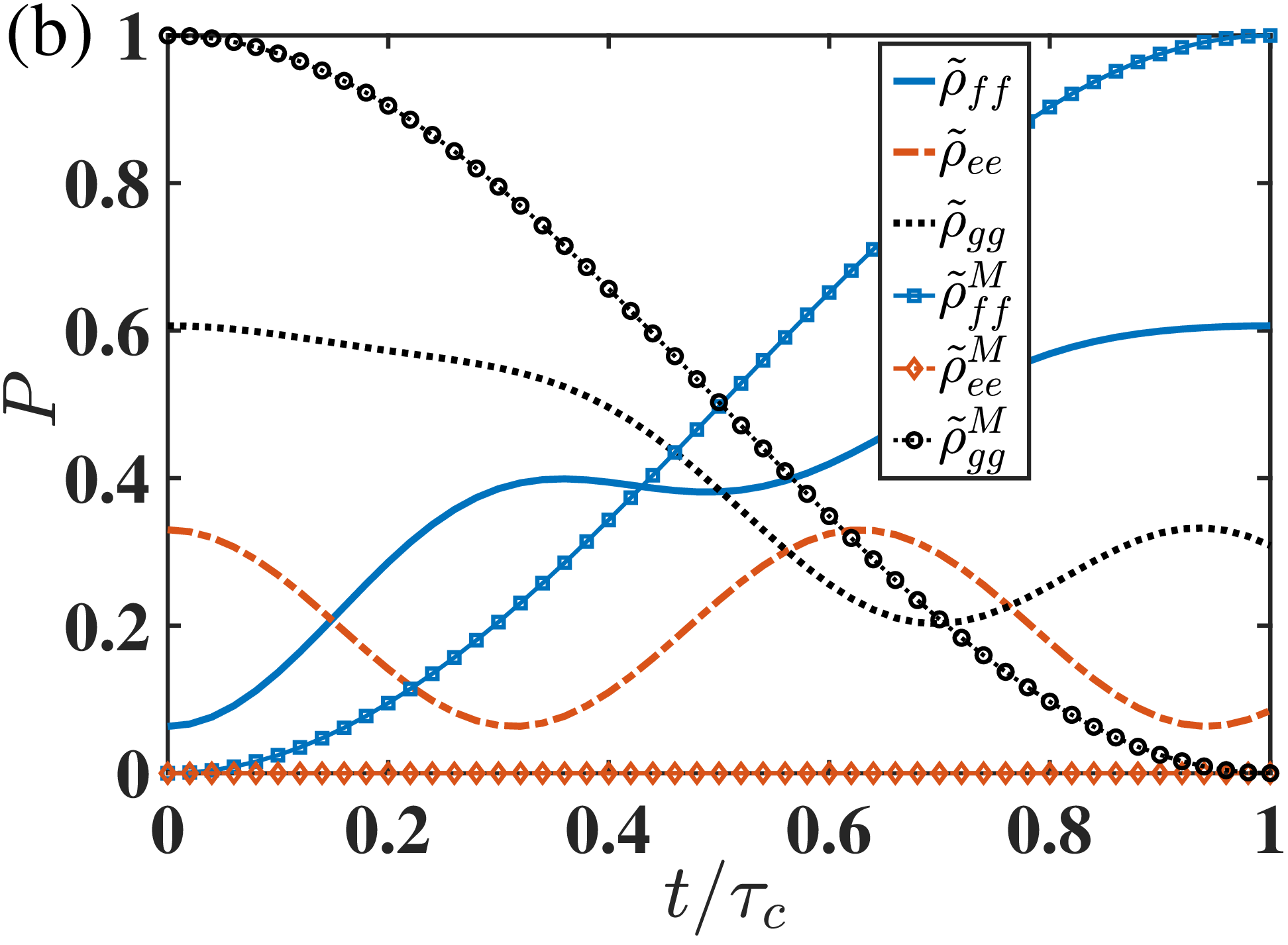}
\caption{Populations over states $|f\rangle$, $|e\rangle$, and $|g\rangle$ are plotted with the dark-dotted lines, the red-dashed lines, and the blue-solid lines, respectively. The lines without markers represent the transition $\textcircled{\scriptsize{3}}\rightarrow\textcircled{\scriptsize{2}}$ from the initial state $\tilde{\rho}(0)$ in Eq.~(\ref{initialstate}) resulting from a self-discharging process with $\gamma_f\tau=0.5$. The lines with markers represent the transition $\textcircled{\scriptsize{3}}\rightarrow\textcircled{\scriptsize{4}}\rightarrow\textcircled{\scriptsize{1}}$ from the initial state $\tilde{\rho}^M(0)$ in Eq.~(\ref{densitymea}). In (a) and (b), the charging periods are set as $\Omega\tau_c=\pi$ and $\Omega\tau_c=5$, respectively. }\label{charge}
\end{figure}
In Fig.~\ref{charge}, we present the STA recharging dynamics in QB populations along different transition paths. For the sake of simplicity and experimental feasibility~\cite{batteryexp}, we apply the sine-wave pulses to both driving pulses,
\begin{equation}\label{sinepulse}
\Omega_1(t)=\Omega\sin\left(\frac{\pi t}{2\tau_c}\right),\quad \Omega_2(t)=\Omega\cos\left(\frac{\pi t}{2\tau_c}\right).
\end{equation}
Then by Eq.~(\ref{omegacd}), we have $\Omega_{\rm CD}(t)=\frac{\pi}{2\tau_c}$.

The lines with no markers in both Figs.~\ref{charge}(a) and (b) indicate the population dynamics from stage $\textcircled{\scriptsize{3}}$ to stage $\textcircled{\scriptsize{2}}$, where the QB has an amount of unextractable energy. It is found that when $\Omega\tau_c=\pi$, that follows the phase condition in Eq.~(\ref{ergotropymax}), the populations over the levels $|g\rangle$ and $|f\rangle$ are mutually exchanged at the end of recharging, i.e., $\tilde{\rho}_{ff}(\tau_c)=\tilde{\rho}_{gg}(0)$ and $\tilde{\rho}_{gg}(\tau_c)=\tilde{\rho}_{ff}(0)$ [see Fig.~\ref{charge}(a)]. The battery system thus goes back to its previous state before work extraction. In Fig.~\ref{charge}(b) with a different phase condition $\Omega\tau_c=5$, it is found that $\tilde{\rho}_{ff}(\tau_c)=\tilde{\rho}_{gg}(0)$, and however, the final ergotropy is much smaller than Eq.~(\ref{ergotropymax}) since $\tilde{\rho}_{gg}(\tau_c)>\tilde{\rho}_{ee}(\tau_c)$ by Eq.~(\ref{ergotro}). In contrast, when the postselection over the ground state $|g\rangle$ is successfully performed, the QB can be fully charged with the maximum ergotropy $\omega_f$ in the end and the final state is insensitive to the choice of the recharging period $\tau_c$ [see the marked lines in both Figs.~\ref{charge}(a) and (b)].

\begin{figure}[htbp]
\centering
\includegraphics[width=0.45\textwidth]{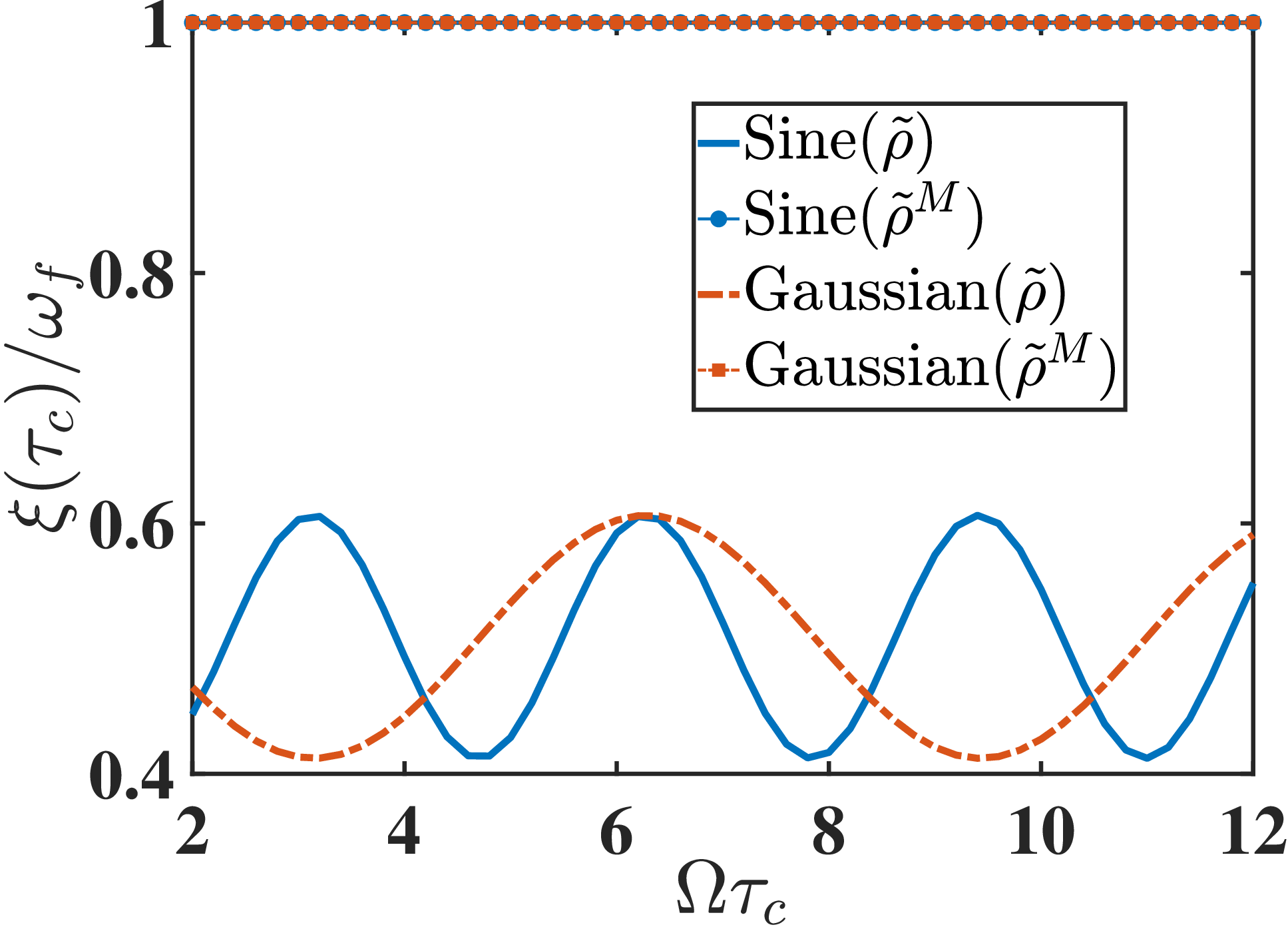}
\caption{Ergotropy $\xi(\tau_c)$ versus the recharging period $\tau_c$ with or without the postselection $M_g$ under sine-wave or Gaussian pulses. The transition frequencies are set as $\omega_f=1.7\omega_e$.}\label{ergotropy}
\end{figure}
Our protocol adopts various shapes of the driving pulses $\Omega_{1,2}(t)$. In Fig.~\ref{ergotropy}, we apply both sine-wave and Gaussian pulses to the recharging process as two popular pulses in existing works for STIRAP~\cite{STIRAP,STIRAP2}. The Gaussian pulses can be described as
\begin{equation}\label{Gausspulse}
\begin{aligned}
\Omega_1(t)&=\Omega\exp\left[-\frac{(t-\tau_c-\alpha)^2}{\sigma^2}\right],\\
\Omega_2(t)&=\Omega\exp\left[-\frac{(t-\tau_c+\alpha)^2}{\sigma^2}\right].
\end{aligned}
\end{equation}
One can then explicitly find the pulse for the CD term
\begin{equation}\label{GausspulseCD}
\Omega_{\rm CD}(t)=\frac{2\alpha}{\sigma^2}\cosh^{-1}\left(\frac{4\alpha t-2\alpha\tau_c}{\sigma^2}\right)
\end{equation}
according to Eq.~(\ref{omegacd}). In numerical simulations, the pulse parameters are set as $\alpha=\tau_c/10$ and $\sigma=\tau_c/6$ to approximately meet the boundary conditions for the adiabatic passage of the dark state $|\lambda_0(t)\rangle$.

Figure~\ref{ergotropy} demonstrates the distinct ergotropy $\xi(\tau_c)$ under the recharging protocols with and without postselection by $M_g$. It is found that along the measurement-free path $\textcircled{\scriptsize{3}}\rightarrow \textcircled{\scriptsize{2}}$, $\xi(\tau_c)$ can attain periodically its maximal value $\xi_{\rm max}(\tau_c)$ in Eq.~(\ref{ergotropymax}) for either sine-wave or Gaussian pulses (see the blue solid line and the red dashed line with no markers). The latter is longer than the former in period. Along the path $\textcircled{\scriptsize{3}}\rightarrow \textcircled{\scriptsize{4}}\rightarrow \textcircled{\scriptsize{1}}$, the initial state of QB becomes $\tilde{\rho}^M(0)$ in Eq.~(\ref{densitymea}) under the postselection instead of $\tilde{\rho}(0)$ in Eq.~(\ref{initialstate}). Therefore, the ergotropy $\xi(\tau_c)$ remains $\omega_f$, regardless of the shape of the driving pulses (see the blue solid line marked with circles and the red dashed line marked with squares).

\section{Systematic errors and decoherence on charging}\label{Sec:syserror}
In the ideal situation, our recharging protocol assisted by the STA method in Sec.~\ref{Sec:charging} is based on the adiabatic trajectory of the dark state $|\lambda_0(t)\rangle$ in Eq.~(\ref{eigenstr}). In practice, the control over the varying parameters is however not exactly implemented because of technical imperfections and constraints. Moreover, environmental decoherence can induce self-discharging in the recharging process since the quantum battery is inevitably an open system. In this section, we investigate the effects of systematic errors and decoherence on the charging performance with respect to the battery ergotropy. In the presence of errors or noises, the final state may not satisfy the conditions $\rho_{ff}\ge\rho_{ee}\ge\rho_{gg}$ and $\rho_{fe}=\rho_{fg}=\rho_{eg}=0$. The unitary transformation that completely extracts the QB energy thus will deviate from $U_{\rm w}$~\eqref{Uwork}. In the following numerical evaluation, the ergotropy is evaluated by its definition in Eq.~(\ref{ergoint2}) and the initial state is fixed as $\tilde{\rho}^M(0)$ in Eq.~(\ref{densitymea}).

\subsection{Systematic errors on driving pulses}\label{Sec:drivingerror}
\begin{figure}[htbp]
\centering
\includegraphics[width=0.45\textwidth]{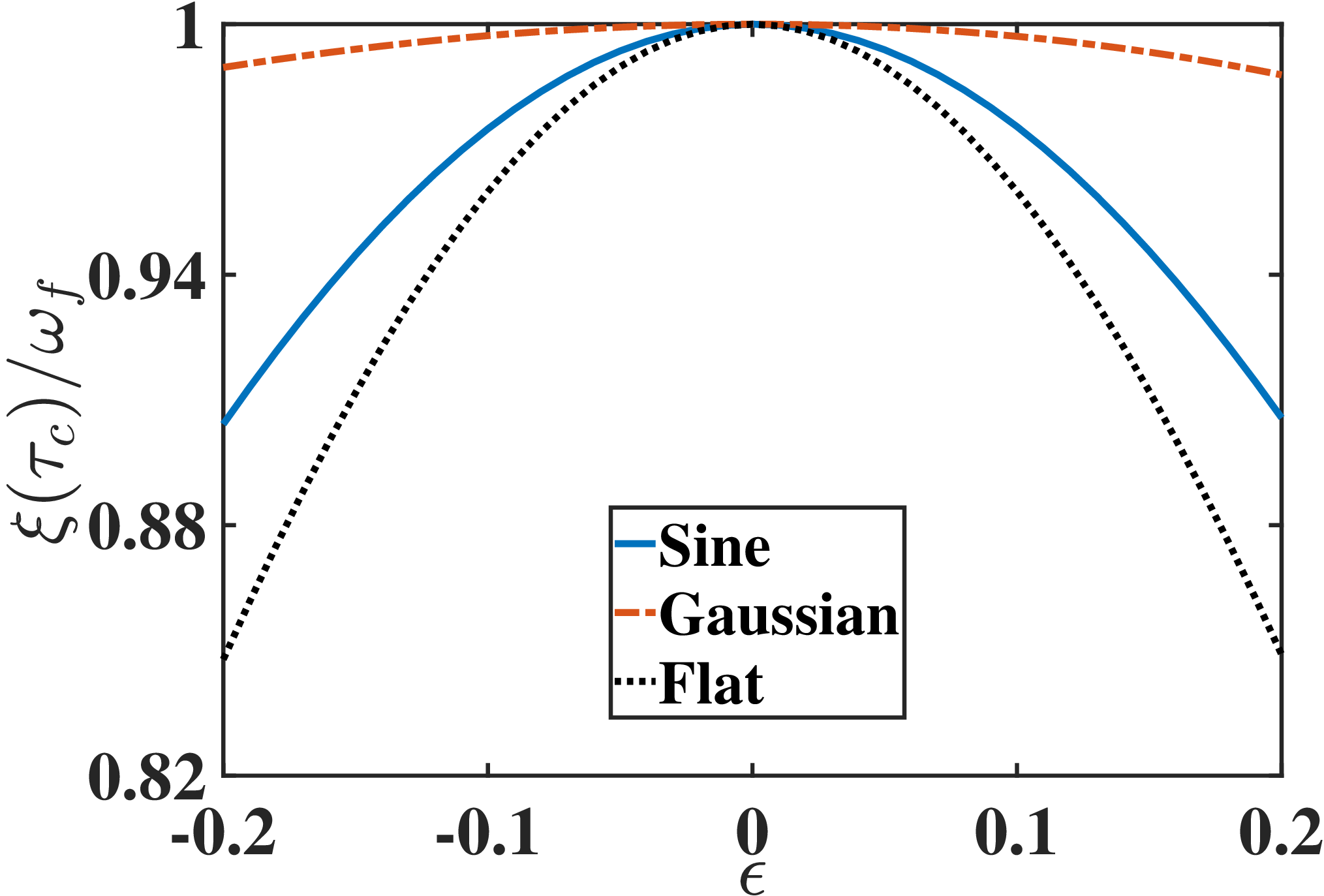}
\caption{Final ergotropy $\xi(\tau_c)$ as a function of the intensity error $\epsilon$ under various driving pulse shapes. The transition frequencies $\omega_f=1.7\omega_e$ and the recharging period $\Omega\tau_c=\pi$.}\label{syserror}
\end{figure}
We first consider the systematic deviation in the driving intensities of the pulses. In particular, we suppose that in experiments the full STA Hamiltonian~(\ref{Hsta}) becomes
\begin{equation}\label{Hstaexprabi}
\begin{aligned}
H_{\rm exp}&=\Omega_1(t)(1+\epsilon)|g\rangle\langle e|+\Omega_2(t)(1-\epsilon)|e\rangle\langle f|\\
&+i\Omega_{\rm CD}(t)|g\rangle\langle f|+{\rm H.c.},
\end{aligned}
\end{equation}
where $\epsilon$ is a dimensionless coefficient implying the relative deviation on $\Omega$.

In Fig.~\ref{syserror}, we compare the error sensitivities of the driving intensities under various driving-pulse shapes when $\Omega\tau_c=\pi$, including the sine-wave pulses (the blue solid line), the Gaussian pulses (the red dashed line), and the flat pulses (the dark dotted line). Flat means that the pulses are square-wave functions of time lasting $\tau_c$, whose magnitudes are $\Omega_1=\Omega_2=\Omega/\sqrt{2}$. It turns out to be a passage with $\Omega_{\rm CD}=0$. It is found that the ergotropy $\xi(\tau_c)$ generated by recharging with the Gaussian pulses demonstrates a much stronger robustness than the sine-wave pulses and the flat pulses. In particular, the ergotropy can be maintained as large as $\xi(\tau_c)\ge0.98\omega_f$ in the range of the normalized error $-0.2\le\epsilon\le0.2$. With the flat pulses, the QB ergotropy declines to $0.85\omega_f$ when $|\epsilon|=0.2$.
\begin{figure}[htbp]
\centering
\includegraphics[width=0.23\textwidth]{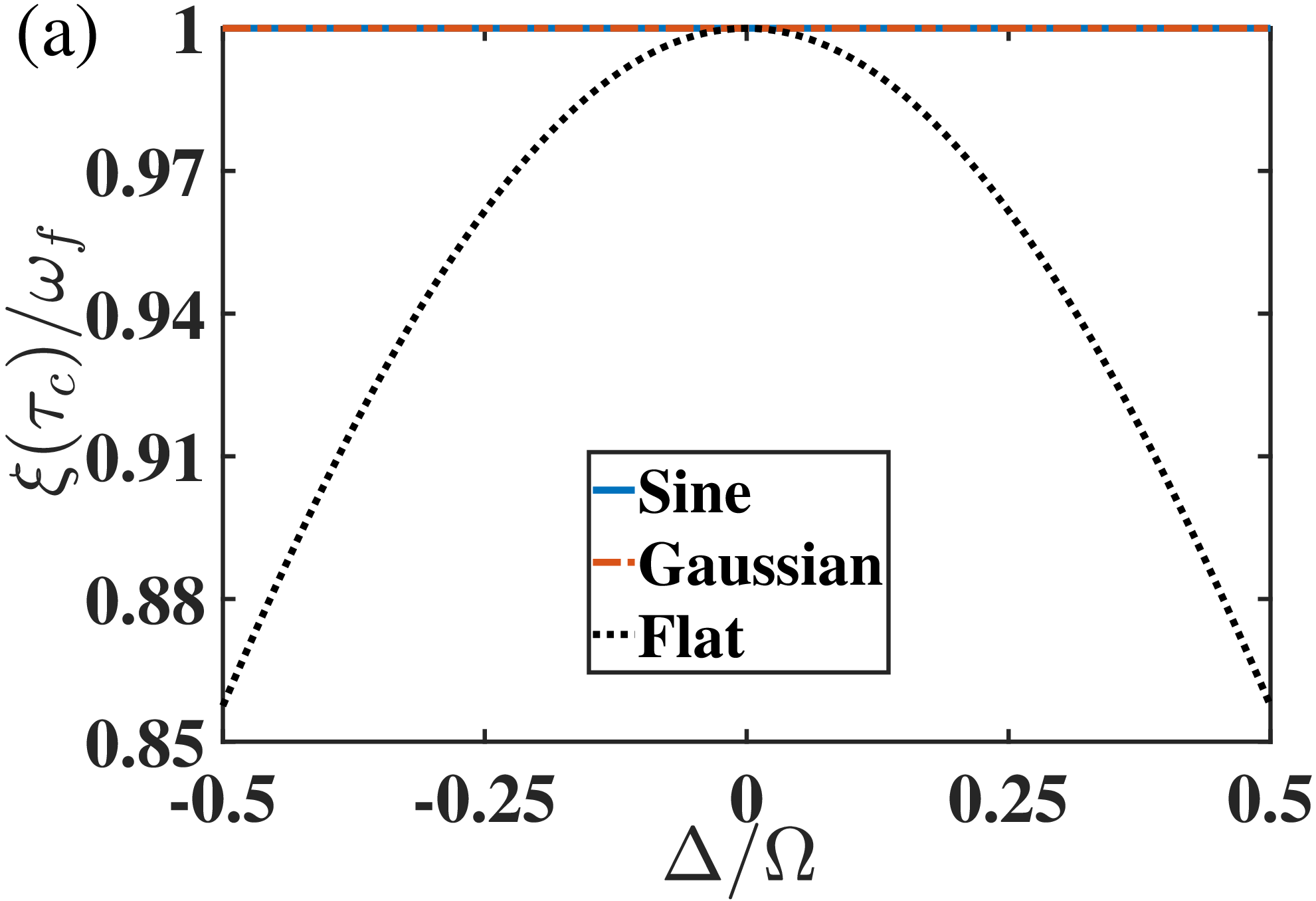}
\includegraphics[width=0.23\textwidth]{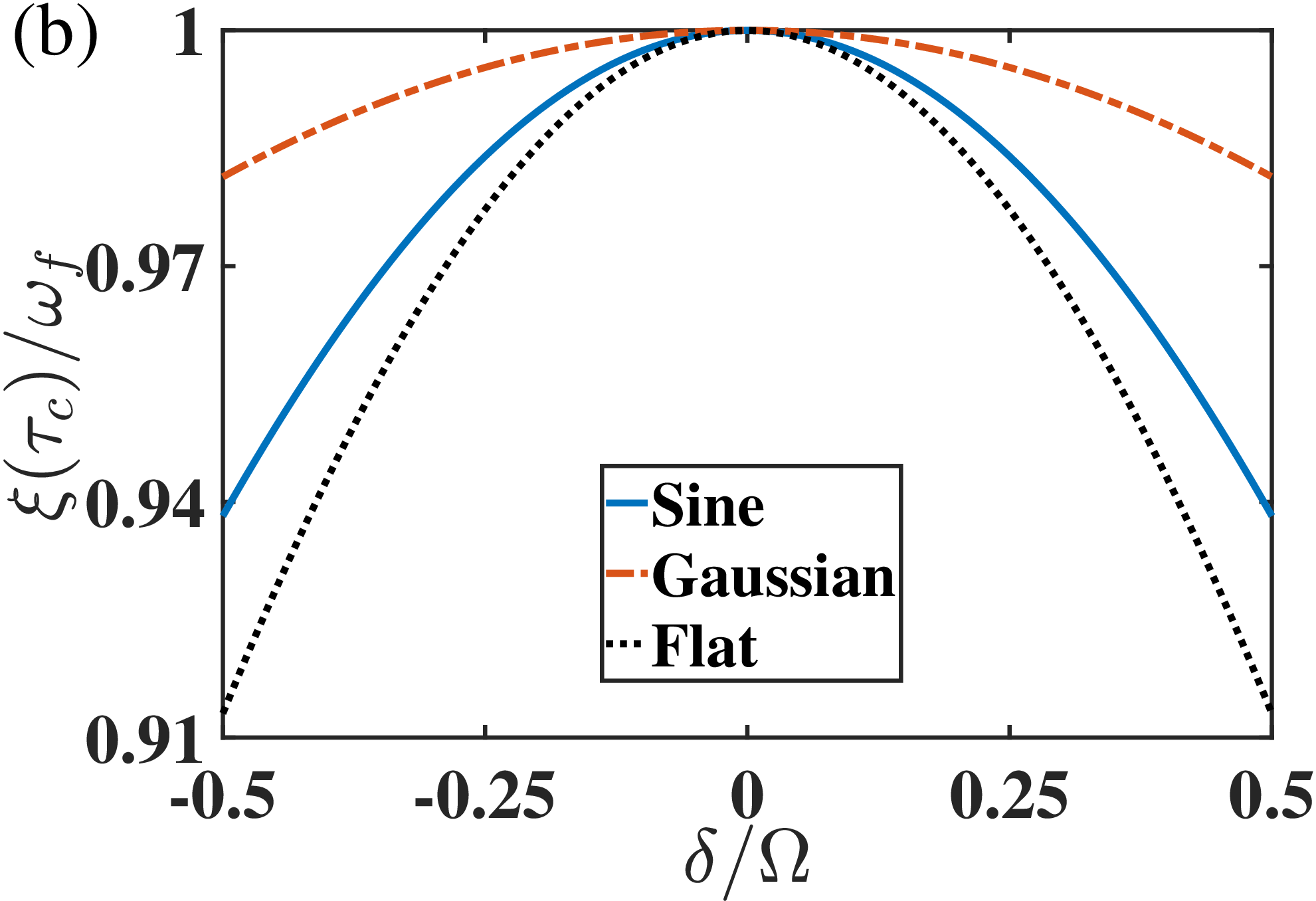}
\caption{Final ergotropy $\xi(\tau_c)$ as a function of the systematic errors associated with the driving frequency derivations (a) $\Delta$ and (b) $\delta$ under various driving pulse shapes. In (a), $\delta=0$, and in (b), $\Delta=0$. Here the parameters $\omega_f=1.7\omega_e$  and  $\Omega\tau_c=\pi$.}\label{syserrordetun}
\end{figure}

Then we consider the sensitivity of the recharging protocol to the deviations of the driving frequencies $\omega_1$ and $\omega_2$ in Eq.~(\ref{drivHam}). In this case, we have
\begin{equation}\label{Hstaexpfre}
\begin{aligned}
H'_{\rm exp}&=\Delta|e\rangle\langle e|+\delta|f\rangle\langle f|+\big[\Omega_1(t)|g\rangle\langle e|
+\Omega_2(t)|e\rangle\langle f|\\
&+i\Omega_{\rm CD}(t)|g\rangle\langle f|+{\rm H.c.}\big],
\end{aligned}
\end{equation}
where $\Delta\equiv\omega_e-\omega_1$ and $\delta\equiv\omega_f-\omega_1-\omega_2$ are the detunings between the driving frequencies and the qutrit transition frequencies $\omega_{e,f}$.

The recharging via the adiabatic path of $|\lambda_0(t)\rangle$ is independent of the detuning $\Delta$. Then one can expect that the STA recharging with arbitrary shapes of pulses is insensitive to $\Delta$, as shown in Fig.~\ref{syserrordetun}(a). In the range of $-0.5<\Delta/\Omega<0.5$, the ergotropy can be maintained nearly $\omega_f$ for both sine-wave and Gaussian pulses. While it drops to about $0.86\omega_f$ for the flat pulse when $|\Delta/\Omega|=0.5$. Figure~\ref{syserrordetun}(b) demonstrates the ergotropy in the presence of the detuning associated with the state $|f\rangle$, which is relevant to the dark state. Still, the ergotropy of QB charged by the Gaussian pulses exhibits a stronger robustness than the sine-wave pulses. In the range of $-0.5<\delta/\Omega<0.5$, we have $\xi(\tau_c)\ge0.98\omega_f$ for the Gaussian shape and $\xi(\tau_c)\ge0.93\omega_f$ for the sine-wave shape. The flat pulses yield the most fragile charging protocol.

\subsection{External decoherence}\label{Sec:decoherence}
\begin{figure}[htbp]
\centering
\includegraphics[width=0.23\textwidth]{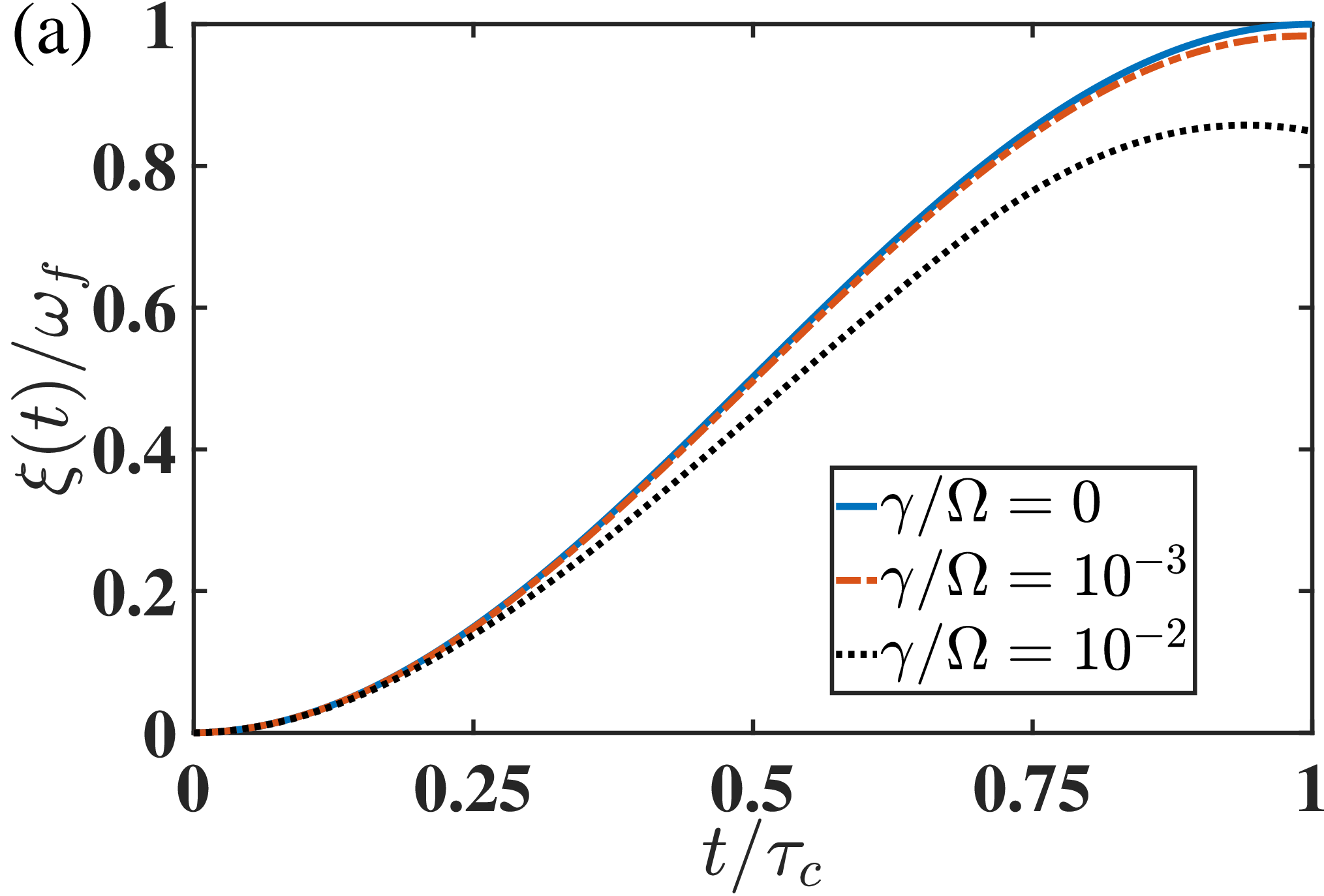}
\includegraphics[width=0.23\textwidth]{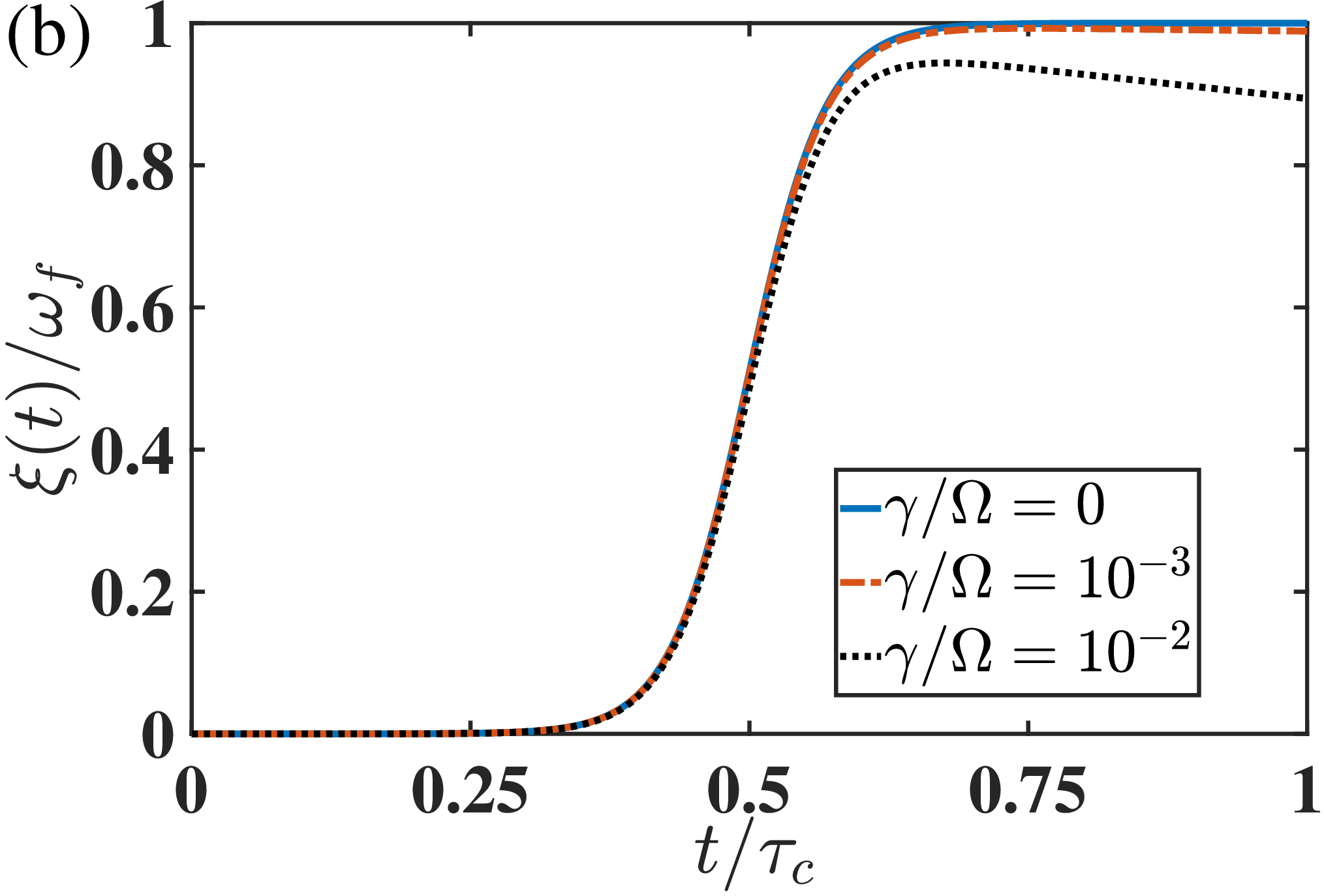}
\caption{Ergotropy dynamics in the presence of environmental decoherence under the charging with (a) sine-wave pulses and (b) Gaussian pulses. The decoherence rates in Eq.~(\ref{lindblad}) are set the same as Fig.~\ref{selfdischarge}. Here the parameters $\omega_f=1.7\omega_e$  and  $\Omega\tau_c=\pi$.}\label{chargedecoherence}
\end{figure}
In this section, we take the self-discharging by decoherence during the adiabatic recharging into account. The recharging dynamics of QB is then governed by the Lindblad master equation~(\ref{lindblad}), where the bare Hamiltonian $H_0$ is replaced with the STA Hamiltonian $H_{\rm STA}$ in Eq.~(\ref{Hsta}).

Figures~\ref{chargedecoherence}(a) and (b) demonstrate the dynamics of the QB ergotropy under charging with sine-wave and Gaussian pulses, respectively. Here the decoherence rates characterized with $\gamma$ are set the same as Fig.~\ref{selfdischarge}. The dynamical behaviors are dependent on the shapes of pulses. For the sine-wave pulse, the ergotropy increases almost with the same rate until approaches almost unit when $t/\tau_c\rightarrow1$. It is found that $\xi(\tau_c)\leq0.98\omega_f$ when $\gamma/\Omega\le 10^{-3}$ and $\xi(\tau_c)$ drops to about $0.85\omega_f$ when $\gamma$ is as large as $10^{-2}\Omega$. For the Gaussian pulses, the ergotropy can be maintained above $0.99\omega_f$ when $\gamma/\Omega\le 10^{-3}$. The ergotropy declines to $0.89\omega_f$ when $\gamma/\Omega=10^{-2}$. Comparing Fig.~\ref{chargedecoherence}(a) with Fig.~\ref{chargedecoherence}(b), one can observe that the ergotropy of Gaussian pulses is higher than that of the sine-wave pulses with the same decay rate. The charging protocol using Gaussian pulses is more robust against environmental noise than that using the sine-wave pulses. Under the Gaussian pulses, the QB is almost in the ground state before the charging process starting from about $0.3\tau_c$, so that the cumulated influence from the environmental noise is less than that under the sine-wave pulses.

\section{Discussion}\label{Sec:dis}
\subsection{Physical implementation}\label{Sec:physic}
Our recharging protocol using the STA method can be implemented in various experimental platforms, including the superconducting circuit~\cite{superqubit,superqubit2}, the trapped ion~\cite{trapion}, and the Rydberg atom~\cite{rydberg}. If the $\Xi$-type qutrit in Fig.~\ref{diagram}(a) does not allow to pump a microwave pulse to the transition between $|g\rangle$ and $|f\rangle$ under the selection rule, one can then implement the CD Hamiltonian by applying a two-photon process. It is generated by an extra driving field with frequency $\omega_p=\omega_f/2$ coupled to the transitions $|g\rangle\leftrightarrow|e\rangle$ and $|e\rangle\leftrightarrow|f\rangle$ with the Rabi frequencies $\Omega_p$ and $\sqrt{2}\Omega_p$, respectively~\cite{staatom2}. In particular, the driving Hamiltonian can be written as
\begin{equation}\label{Hp}
H_p=\Omega_p(t)e^{i\phi+i\omega_pt}\left(|g\rangle\langle e|+\sqrt{2}|e\rangle\langle f|\right)+{\rm H.c.}.
\end{equation}
An effective coupling between $|g\rangle$ and $|f\rangle$ arises in the dispersive regime $\delta_d=\omega_e-\omega_p\gg\Omega_p$ with $\phi=\pi/4$. In this case, we have
\begin{equation}\label{Heff}
H_{\rm eff}=i\Omega_{\rm eff}(t)|g\rangle\langle f|+{\rm H.c.},
\end{equation}
where $\Omega_{\rm eff}(t)=\sqrt{2}\Omega^2_p(t)/\delta_d$. Then by setting $\Omega_{\rm eff}(t)=\Omega_{\rm CD}(t)$, the demanded CD term in Eq.~(\ref{Hsta}) can be indirectly realized.

In the superconducting circuit, the QB can be set up in a $\Delta$-type flux qutrit~\cite{deltaqubit}. It allows all three dipole transitions among $|g\rangle$, $|e\rangle$, and $|f\rangle$ when $\Phi/\Phi_0\ne 0.5$, indicating no forbidden transition. $\Phi$ is the static magnetic flux through the loop and $\Phi_0$ is the magnetic-flux quantum. The counterdiabatic driving term can thus be directly performed between $|g\rangle$ and $|f\rangle$.

\subsection{Charging energetic cost}\label{Sec:ener}
The energetic cost~\cite{energycost,energycost2} to implement the unitary operation $U(t)$ in Eq.~(\ref{timeUspace}) for QB can be given by
\begin{equation}\label{cost}
C\equiv\frac{1}{\tau_c}\int^{\tau_c}_0||H_{\rm STA}(t)||dt,
\end{equation}
where $||H_{\rm STA}(t)||=\sqrt{{\rm Tr}[H_{\rm STA}^2(t)]}$ is the Hilbert-Schmidt norm of the full Hamiltonian in Eq.~(\ref{Hsta}) for our recharging protocol with the transitionless driving. Consequently, we have
\begin{equation}\label{coststa}
C=\frac{\sqrt{2}}{\tau_c}\int^{\tau_c}_0\sqrt{\Omega^2_1(t)+\Omega^2_2(t)+\Omega^2_{\rm CD}(t)}dt.
\end{equation}
A charging efficiency $\eta$ can then be defined as the ratio of the ergotropy variation and the total energy $E_{\rm tot}$ consumed in the battery recharging for STA evolution $\textcircled{\scriptsize{3}}\rightarrow\textcircled{\scriptsize{2}}$~\cite{batteryexp,chargingefficiency}, 
\begin{equation}\label{efficiency}
\eta\equiv\frac{\xi_{\rm max}(\tau_c)}{E_{\rm tot}}=\frac{\xi_{\rm max}(\tau_c)}{\xi_{\rm max}(\tau_c)+C}.
\end{equation}

The projective measurement is accompanied by a change of information~\cite{measurecost1,measurecost2,measurecost3}, i.e., by a change of the von Neumann entropy of the system, that will cost an amount of energy
\begin{equation}\label{meacost}
\begin{aligned}
C_M=k_BT\left({\rm Tr}\{\tilde{\rho}(0)\ln[\tilde{\rho}(0)]\}-{\rm Tr}\{\tilde{\rho}^M(0)\ln[\tilde{\rho}^M(0)]\}\right),
\end{aligned}
\end{equation}
where $k_B$ is the Boltzmann constant and $T$ is the temperature of the measurement device. As an ideal low-bound, $C_M$ is actually an approximated result when the measurement device uses quantum resources, such as single-photon detection. While when the device uses classical resources, such as coherent states, the energetic cost will become much greater~\cite{measurecost2}. Nevertheless, the charging efficiency~\cite{batteryexp,chargingefficiency} for the whole transition $\textcircled{\scriptsize{3}}\rightarrow\textcircled{\scriptsize{4}}\rightarrow\textcircled{\scriptsize{1}}$ can be expressed as
\begin{equation}\label{efficiencytot}
\eta=\frac{\omega_f}{E_{\rm tot}}=\frac{\omega_f}{\omega_f+C+C_M}.
\end{equation}

The energy cost $C$ for the sine-wave pulses can be obtained as $C=\sqrt{8\Omega^2\tau^2_c+\pi^2}/(2\tau_c)$ by Eq.~(\ref{cost}). In recent superconducting qutrit systems~\cite{staatom2}, the transition frequency is about $\omega_f\sim 10$GHz, the Rabi frequency is $\Omega\sim 10-100$MHz, and the period is $\tau_c\sim 100$ns. Take $\Omega=0.001\omega_f$ and $\Omega\tau_c=\pi$ as an example, we have $C=3\Omega/2$. According to Eq.~(\ref{efficiency}), the efficiency for the direct recharging process $\textcircled{\scriptsize{3}}\rightarrow\textcircled{\scriptsize{2}}$ is about $\eta\approx99.7\%$. Suppose that the battery is placed in a low-temperature environment with $T=10$ mK, the ideal energetic cost for postselection is about $C_M\approx0.017\omega_f$ by Eq.~(\ref{meacost}). Then the efficiency given in Eq.~(\ref{efficiencytot}) for the recharging process with a postselection is about $\eta\approx98.2\%$.

\section{Conclusion}\label{Sec:conclusion}
This work focuses on the reusability of the quantum battery. We propose a fast and stable recharging protocol for a three-level quantum battery after it has experienced a period of self-discharging and an amount of work extraction. In contrast to many existing quantum charging protocols, the initial state of our protocol is a passive state characterized by unextractable energy. Our recharging protocol is based on the instantaneous dark state of the battery system that is determined by the STIRAP driving assisted with extra counterdiabatic driving. To avoid the defect that the battery returns only to the state before the work extraction by the charging pulses, we apply a postselection with a projective measurement before charging to refresh a full-ergotropy battery. And the postselection does not have a significant impact on the energy cost in charging.

Our protocol is found to be robust against the systematic errors arising from the deviations of microwave driving intensities and driving frequencies. Moreover, the recharging performance of our battery is resilient to both energy dissipation and quantum dephasing. In practice, the counterdiabatic driving in our recharging protocol can be effectively realized in a three-level atomic system with the forbidden transition. In the case of parallel charging with individual environments, our protocol is scalable for the $N$-atomic system. Our findings thus promise a remarkable promotion for quantum battery, which is also an interesting application of shortcut to adiabaticity.

\section*{Acknowledgements}
We acknowledge financial support from the National Natural Science Foundation of China (Grants No. 12404405 and Grant No. 11974311) and the Science Foundation of Hebei Normal University of China (Grant No. 13114122).

\appendix

\section{STA versus STIRAP in charging}\label{appa}
We compare the charging performance of the protocols based on STIRAP and STA with the extra counterdiabatic driving in terms of population dynamics at a fixed charging period $\tau_c$ and the final ergotropy under varying $\tau_c$. Here the initial state of QB is $\tilde{\rho}^M(0)=|g\rangle\langle g|$, i.e., stage $\textcircled{\scriptsize{4}}$ under a desired postselection. We consider the evolution in the closed-system scenario.

For the STIRAP Hamiltonian used in previous charging protocols~\cite{stablebattery,batteryexp}, the battery evolution is driven by the Hamiltonian in Eq.~(\ref{Ht}), i.e.,
\begin{equation}\label{Htappendix}
H(t)=\Omega_1(t)|g\rangle\langle e|+\Omega_2(t)|e\rangle\langle f|+\rm{H.c.}.
\end{equation}
A full charging demands a sufficiently long charging period $\tau_c$ under the constraint of the adiabatic approximation. Otherwise, the system could not remain at the dark state $|\lambda_0\rangle$ in Eq.~(\ref{eigenstr}), and its transition to the other eigenstates becomes inevitable. Consequently, the ergotropy of QB cannot attain its maximum value as a result of a nonvanishing population on the middle level $|e\rangle$. An extra CD term in Eq.~(\ref{Hcd}) can suppress the unwanted transitions between instantaneous eigenstates. Thus, we can use the Hamiltonian in Eq.~(\ref{Hsta}), i.e.,
\begin{equation}\label{Hstaappendix}
H_{\rm STA}(t)=\Omega_1(t)|g\rangle\langle e|+\Omega_2(t)|e\rangle\langle f|+i\Omega_{\rm CD}(t)|g\rangle\langle f|+\rm{H.c.}
\end{equation}
to achieve a perfect adiabatic dynamic. Using the STA charging protocol assisted by the CD driving, the battery system can follow the desired adiabatic path $|\lambda_0\rangle$ within a much reduced period $\tau_c$.

\begin{figure}[htbp]
\centering
\includegraphics[width=0.45\textwidth]{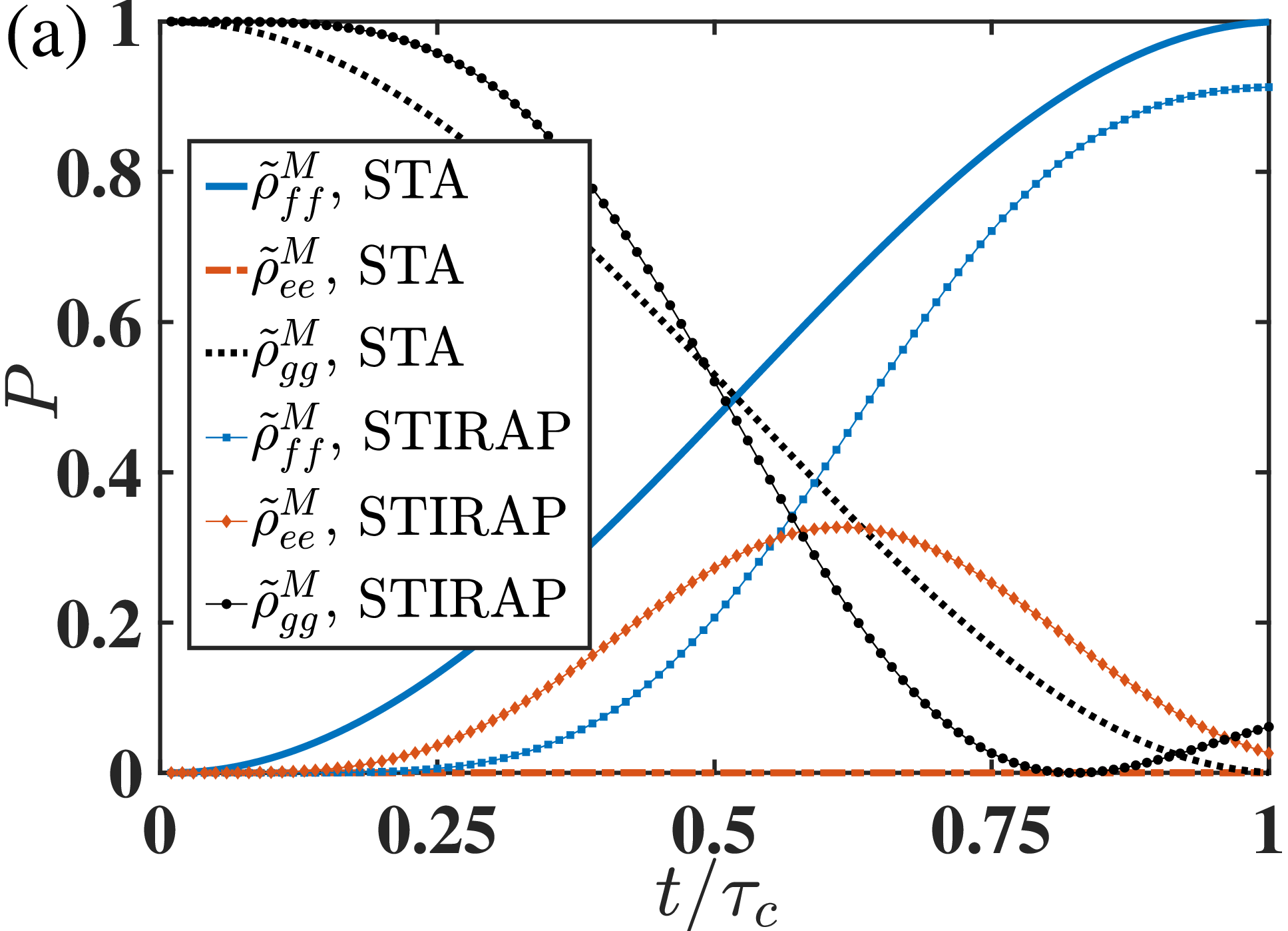}
\includegraphics[width=0.45\textwidth]{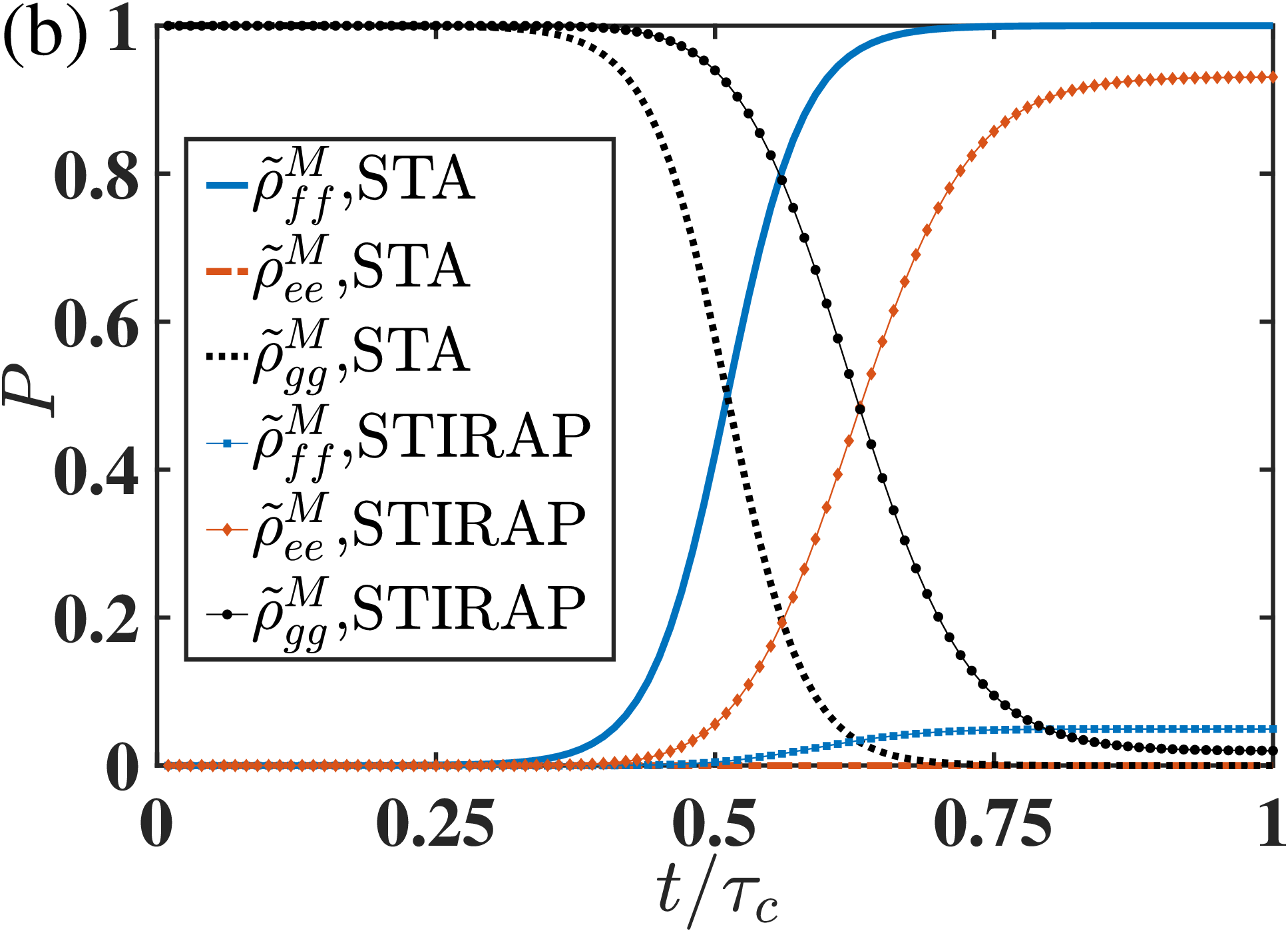}
\caption{Population dynamics of a three-level QB under the STA protocol with $H_{\rm STA}(t)$ in Eq.~(\ref{Hsta}) or the STIRAP protocol with $H(t)$ in Eq.~(\ref{Ht}). The recharging period is fixed as $\Omega\tau_c=5$. In (a) and (b), the driving pulses are the sine-wave and the Gaussian types, respectively. }\label{STAcharge}
\end{figure}

In Fig.~\ref{STAcharge}, it is found that the QB can be completely transformed from the initial ground state $|g\rangle$ to the full-ergotropy state $|f\rangle$ under the STA protocol with either sine-wave or Gaussian pulses [see the blue solid lines]. In sharp contrast, the middle level is significantly populated under the STIRAP protocol and the final population on $|f\rangle$ is about $0.9$ with the sine-wave pulses [see the blue solid line with squares in Fig.~\ref{STAcharge}(a)] and less than $0.1$ with the Gaussian pulses [see the blue solid line with squares in Fig.~\ref{STAcharge}(b)]. Clearly, the conventional STIRAP protocol fails to quickly charge the QB.

\begin{figure}[htbp]
\centering
\includegraphics[width=0.45\textwidth]{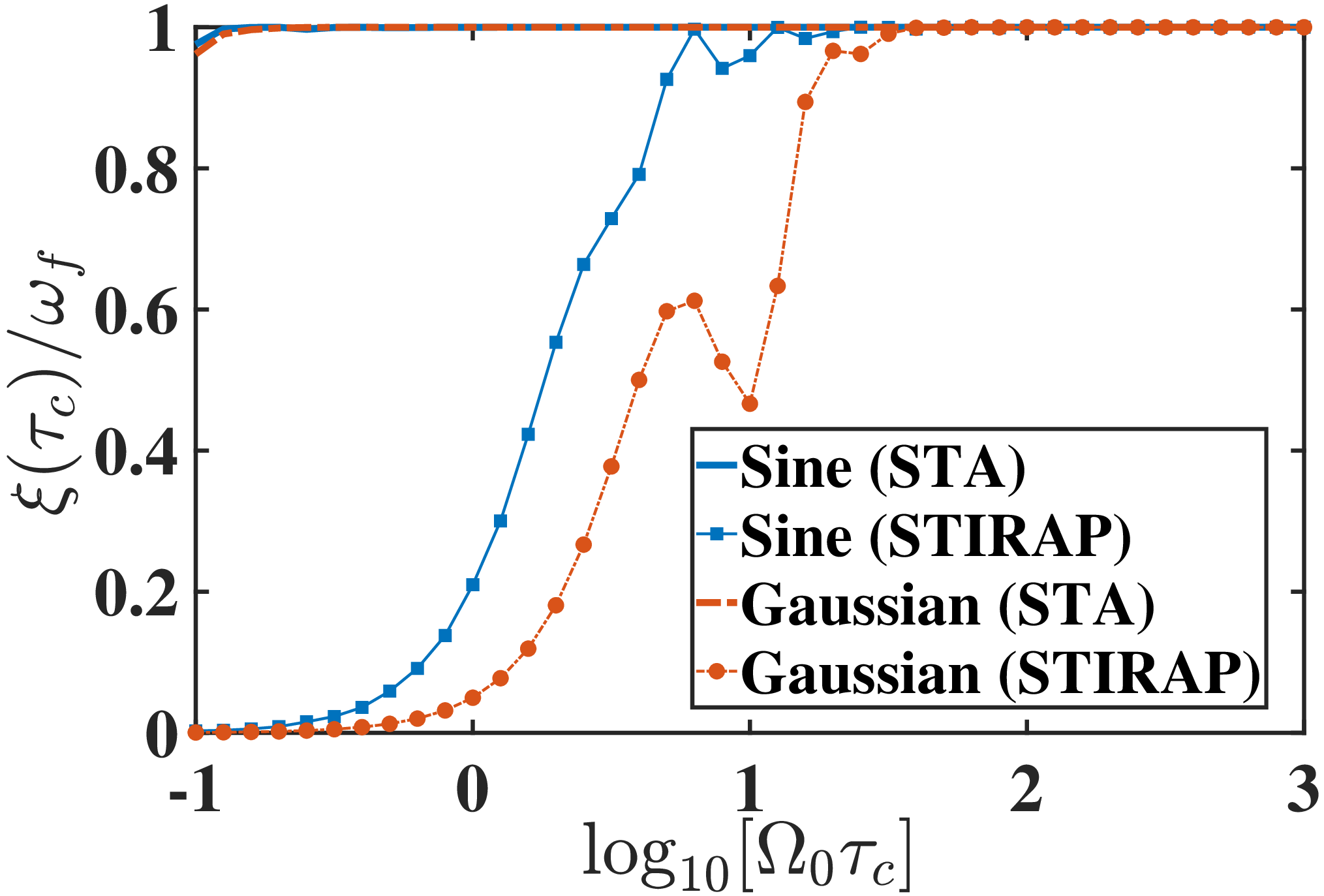}
\caption{Final ergotropy as a function of the recharging period $\tau_c$ under various protocols and driving pulses. $\omega_f=1.7\omega_e$. }\label{STIRAP}
\end{figure}
More explicitly, the final ergotropy $\xi(\tau_c)$ at the end of recharging in Fig.~\ref{STIRAP} demonstrates the acceleration advantage of our STA charging protocol over the conventional STIRAP protocol. It is found that the STA charging protocol can give rise to the maximum ergotropy even if the charging period is as short as $\Omega\tau_c\approx0.15$. In contrast, the ergotropy under the STIRAP protocol is dramatically lower than that under the STA protocol until the adiabatic limit, which is about $\Omega\tau_c\approx30$. In addition, the sine-wave pulse is superior to the Gaussian pulse before the ergotropy is saturated.

\bibliographystyle{elsarticle-num}
\bibliography{reference}

\end{document}